\documentclass[reqno,11pt,a4paper,final]{amsart}
\usepackage[a4paper,left=30mm,right=30mm,top=30mm,bottom=30mm,marginpar=20mm]{geometry} 
\usepackage{amsmath}
\usepackage{amssymb}
\usepackage{amsthm}
\usepackage{amscd}
\usepackage{stmaryrd}
\usepackage{esint}
\usepackage[ansinew]{inputenc}
\usepackage{cite}
\usepackage{bbm}
\usepackage{xcolor}
\usepackage[english=american]{csquotes}
\usepackage[final]{graphicx}
\usepackage{hyperref}
\usepackage{calc}
\usepackage{mathptmx}
\usepackage{bm}
\usepackage{enumerate}
\usepackage[shortlabels]{enumitem}
\usepackage{transparent}
\usepackage[normalem]{ulem}

\graphicspath{{pics/}}

\numberwithin{equation}{section}

\newtheoremstyle{thmlemcorr}{10pt}{10pt}{\itshape}{}{\bfseries}{.}{10pt}{{\thmname{#1}\thmnumber{ #2}\thmnote{ (#3)}}}
\newtheoremstyle{thmlemcorr*}{10pt}{10pt}{\itshape}{}{\bfseries}{.}\newline{{\thmname{#1}\thmnumber{ #2}\thmnote{ (#3)}}}
\newtheoremstyle{remexample}{10pt}{10pt}{}{}{\bfseries}{.}{10pt}{{\thmname{#1}\thmnumber{ #2}\thmnote{ (#3)}}}

\theoremstyle{thmlemcorr}
\newtheorem{theorem}{Theorem}
\numberwithin{theorem}{section}

\theoremstyle{thmlemcorr*}
\newtheorem{theorem*}{Theorem}
\newtheorem{lemma*}[theorem]{Lemma}
\newtheorem{corollary*}[theorem]{Corollary}
\newtheorem{proposition*}[theorem]{Proposition}
\newtheorem{problem*}[theorem]{Problem}
\newtheorem{conjecture*}[theorem]{Conjecture}
\newtheorem{definition*}[theorem]{Definition}

\theoremstyle{remexample}

\newtheorem{example}[theorem]{Example}

\newcommand{\Crm}{\mathrm{C}}

\newcommand{\Erm}{\mathrm{E}}

\newcommand{\Irm}{\mathrm{I}}

\newcommand{\Srm}{\mathrm{S}}
\newcommand{\Trm}{\mathrm{T}}

\newcommand{\Bcal}{\mathcal{B}}

\newcommand{\Dcal}{\mathcal{D}}
\newcommand{\Ecal}{\mathcal{E}}

\newcommand{\Hcal}{\mathcal{H}}
\newcommand{\Ical}{\mathcal{I}}

\newcommand{\Lcal}{\mathcal{L}}
\newcommand{\Mcal}{\mathcal{M}}

\newcommand{\Pcal}{\mathcal{P}}

\newcommand{\Scal}{\mathcal{S}}

\newcommand{\Wcal}{\mathcal{W}}

\newcommand{\Gfrak}{\mathfrak{G}}

\newcommand{\Pfrak}{\mathfrak{P}}

\newcommand{\Zfrak}{\mathfrak{Z}}

\newcommand{\pfrak}{\mathfrak{p}}

\newcommand{\Mbf}{\mathbf{M}}

\DeclareMathOperator{\Id}{Id}

\DeclareMathOperator{\im}{im}

\DeclareMathOperator{\Diverg}{Div}
\DeclareMathOperator{\curl}{curl}

\DeclareMathOperator{\tr}{tr}
\DeclareMathOperator{\spn}{span}

\DeclareMathOperator{\supp}{supp}
\DeclareMathOperator{\cof}{cof}

\DeclareMathOperator{\dev}{dev}
\DeclareMathOperator{\proj}{proj}

\DeclareMathOperator{\infc}{\square}
\DeclareMathOperator{\Wedge}{{\textstyle\bigwedge}}

\newcommand{\ee}{\mathrm{e}}

\newcommand{\set}[2]{\left\{\, #1 \ \ \textup{\textbf{:}}\ \ #2 \,\right\}}

\newcommand{\setb}[2]{\bigl\{\, #1 \ \ \textup{\textbf{:}}\ \ #2 \,\bigr\}}
\newcommand{\setB}[2]{\Bigl\{\, #1 \ \ \textup{\textbf{:}}\ \ #2 \,\Bigr\}}

\newcommand{\norm}[1]{\|#1\|}

\newcommand{\abs}[1]{|#1|}

\newcommand{\absb}[1]{\bigl|#1\bigr|}

\newcommand{\altnorm}[1]{{\left\vert\kern-0.25ex\left\vert\kern-0.25ex\left\vert #1 \right\vert\kern-0.25ex\right\vert\kern-0.25ex\right\vert}}

\newcommand{\spr}[1]{( #1 )}

\newcommand{\sprb}[1]{\bigl( #1 \bigr)}

\newcommand{\dpr}[1]{\langle #1 \rangle}

\newcommand{\dprb}[1]{\bigl\langle #1 \bigr\rangle}

\newcommand{\dprBB}[1]{\biggl\langle #1 \biggr\rangle}

\newcommand{\dbr}[1]{\llbracket #1 \rrbracket}

\newcommand{\cl}[1]{\overline{#1}}
\newcommand{\di}{\mathrm{d}}
\newcommand{\dd}{\;\mathrm{d}}
\newcommand{\DD}{\mathrm{D}}
\newcommand{\N}{\mathbb{N}}
\newcommand{\R}{\mathbb{R}}

\newcommand{\Z}{\mathbb{Z}}
\newcommand{\loc}{\mathrm{loc}}

\newcommand{\conv}{*}
\newcommand{\hodge}{{\star}}

\newcommand{\sbullet}{\begin{picture}(1,1)(-0.5,-2.5)\circle*{2}\end{picture}}
\newcommand{\frarg}{\,\sbullet\,}

\newcommand{\eps}{\epsilon}

\newcommand{\tv}[1]{\norm{#1}}

\newcommand{\term}[1]{\textbf{#1}}

\newcommand{\GL}{\mathrm{GL}}
\newcommand{\SO}{\mathrm{SO}}
\newcommand{\SL}{\mathrm{SL}}
\newcommand{\Diss}{\mathrm{Diss}}

\newcommand{\tbf}{\mathbf{t}}

\newcommand{\pbf}{\mathbf{p}}

\DeclareMathOperator{\Tan}{T}
\DeclareMathOperator{\Lie}{Lie}

\DeclareMathOperator{\Exp}{Exp}

\newcounter{assumption}
\makeatletter
\newcommand{\nextas}[1]{%
   \refstepcounter{assumption}%
   \protected@write \@auxout{}{\string\newlabel{#1}{{(A\theassumption)}{\thepage}{(A\theassumption)}{#1}{}}}%
   \hypertarget{#1}{(A\theassumption)}%
}
\newcommand{\nextasnamed}[2]{%
   \refstepcounter{assumption}%
   \protected@write \@auxout{}{\string\newlabel{#1}{{(#2)}{\thepage}{(#2)}{#1}{}}}%
   \hypertarget{#1}{(#2)}%
}
\makeatother

\def\XXint#1#2#3{{\setbox0=\hbox{$#1{#2#3}{\int}$} 
\vcenter{\hbox{$#2#3$}}\kern-.5\wd0}}

\usepackage{pict2e}
\makeatletter
\DeclareRobustCommand{\intprod}{%
  \mathbin{\mathpalette\int@prod{(0.1,0)(0.9,0)(0.9,0.8)}}}
\DeclareRobustCommand{\restrict}{%
  \mathbin{\mathpalette\int@prod{(0.1,0.8)(0.1,0)(0.9,0)}}}	
\newcommand{\int@prod}[2]{%
  \begingroup
  \sbox\z@{$\m@th#1+$}%
  \setlength\unitlength{\wd\z@}%
  \begin{picture}(1,1)
  \roundcap
  \polyline#2
  \end{picture}%
  \endgroup
}
\makeatother

\renewcommand{\eps}{\varepsilon}
\renewcommand{\epsilon}{\varepsilon}
\renewcommand{\phi}{\varphi}
\renewcommand{\tilde}{\widetilde}
\renewcommand{\hat}{\widehat}

\begin{document}


\title[Elasto-plastic evolution of single crystals]{Elasto-plastic evolution of single crystals\\driven by dislocation flow}


\author{Thomas Hudson}

\author{Filip Rindler}

\address{Mathematics Institute, University of Warwick, Coventry CV4 7AL, United Kingdom.}
\email{F.Rindler@warwick.ac.uk}




\maketitle


\begin{abstract}
This work introduces a model for large-strain, geometrically nonlinear elasto-plastic dynamics in single crystals. The key feature of our model is that the plastic dynamics are entirely driven by the movement of dislocations, that is, $1$-dimensional topological defects in the crystal lattice. It is well known that glide motion of dislocations is the dominant microscopic mechanism for plastic deformation in many crystalline materials, most notably in metals. We propose a novel geometric language, built on the concepts of space-time ``slip trajectories'' and the ``crystal scaffold'' to describe the movement of (discrete) dislocations and to couple this movement to plastic flow. The energetics and dissipation relationships in our model are derived from first principles drawing on the theories of crystal modeling, elasticity, and thermodynamics. The resulting force balances involve a new configurational stress tensor describing the forces acting against slip. In order to place our model into context, we further show that it recovers several laws that were known in special cases before, most notably the equation for the Peach--Koehler force (linearized configurational force) and the fact that the combination of all dislocations yields the curl of the plastic distortion field. Finally, we also include a brief discussion on how a number of other effects, such as hardening, softening, dislocation climb, and coarse-graining, could be incorporated into our model.

\vspace{4pt}

%
%

\noindent\textsc{Date:} \today{}
\end{abstract}


\section{Introduction}

The modeling of large-strain elasto-plasticity poses many challenges and, despite its great practical importance, no fully satisfactory theory has emerged so far. We refer to~\cite{LemaitreChaboche90,Lubliner08,GurtinFriedAnand10book,HanReddy13,Temam18book} for recent expositions (and many historical references) as well as the original works~\cite{Kroner60,GreenNaghdi71,Rice71,Mandel73,NematNasser79,Dafalias87,LubardaLee81,Naghdi90,Zbib93,OrtizRepetto99,Acharya03,Acharya04,Mielke03a,FrancfortMielke06,Stefanelli08,AcharyaTartar11,AroraAcharya20,Acharya21} for some aspects of this vast field. Most of the existing models are phenomenological in nature and do not attempt to explain plastic distortion from its microscopic origins, but merely describe its macroscopic manifestation. However, at least in crystalline materials, the microscopic origins of plasticity are fairly well understood: Essentially all plastic distortion is caused by the movement of dislocations, that is, $1$-dimensional topological defects in the crystal lattice~\cite{AbbaschianReedHill09,HullBacon11book,AndersonHirthLothe17book}.

The description of dislocations in crystals and their motion are considered in numerous works, for instance~\cite{Kondo55,Nye53,BilbyBulloughSmith55,Noll58,KronerSeeger59,Kroner60,Kroner61,Mura63a,Mura63b,Fox66,Wang67,Willis67,AmodeoGhoniem90a,AmodeoGhoniem90b,Kroner01,Acharya03,Acharya04,BulatovCai06book,CaiArsenlisWeinbergerBulatov06,ContiTheil05,HochrainerZaiserGumbsch07,AbbaschianReedHill09,SandfeldEtAl11,HochrainerEtAl14,HehlObukhov07,LazarAnastassiadis08,LazarAnastassiadis09,KupfermanMaor15,AcharyaTartar11,AroraAcharya20,EpsteinKupfermanMaor20,KupfermanOlami20}. Let us mention in particular recent work on the theory of \enquote{field dislocation mechanics} for the evolution of continuously-distributed dislocations, in which Kr\"{o}ner's dislocation density tensor $\alpha$ is coupled to the elasto-plastic behaviour of a material. This theory was developed by Acharya and collaborators~\cite{Acharya01,Acharya03,Acharya04,AcharyaTartar11,AroraAcharya20,Acharya21} and is a continuation of earlier work by Bilby--Bullough--Smith~\cite{BilbyBulloughSmith55}, Fox~\cite{Fox66}, Kondo~\cite{Kondo55}, Kr\"{o}ner~\cite{Kroner01,Kroner60,Kroner61}, Mura~\cite{Mura63a,Mura63b}, Noll~\cite{Noll58}, Nye~\cite{Nye53}, Wang~\cite{Wang67}, and Willis~\cite{Willis67}, among others. We refer the reader to~\cite[Appendix~A]{AroraAcharya20} and~\cite{EpsteinKupfermanMaor20} for further historical references. We note that despite such developments, these models do not currently have formulations in which well-posedness is rigorously guaranteed, particularly in the singular regime where dislocations are considered as discrete defects; we are only aware of~\cite{AcharyaTartar11} where some mathematical observations are made.

It is therefore the goal of the present work to derive a full model of elasto-plastic evolution driven by (discrete) dislocation motion based on a novel geometric language that can be readily translated into a mathematically rigorous framework. The companion works~\cite{Rindler21a?,Rindler21b?} carry out this translation and prove the first existence theorem for solutions to the full evolutionary system in the rate-independent case.

As in several other papers in the recent mathematical literature~\cite{ContiGarroniMassaccesi15,ContiGarroniOrtiz15,ScalaVanGoethem19,KupfermanOlami20}, we describe individual dislocations on a mesoscale as closed $1$-dimensional loops in a $3$-dimensional body occupying the reference (initial) configuration $\Omega \subset \R^3$. For topological reasons, dislocations are always closed loops inside the crystal specimen, or, if the specimen is polycrystalline, inside a grain. Every dislocation line has a Burgers vector associated with it, which is fixed along the whole dislocation line and specifies the plastic displacement effected when the dislocation moves.

For the evolution of such a dislocation system we introduce the notion of a \emph{slip trajectory}, that is, a $2$-dimensional oriented surface $S$ in the space-time cylinder $[0,T] \times \cl{\Omega}$. The dislocation at time $t$ is then given as the \emph{slice} $S|_t$ of $S$ at time $t$, i.e.\ the intersection of $S$ with the plane $\{t\} \times \R^3$, which can be shown to be a collection of oriented lines; see Figure~\ref{fig:SlipTrajectory} for an illustration. The notion of a slip trajectory can be made mathematically rigorous using the theory of integral currents~\cite{KrantzParks08book,Federer69book}, but we present the model with minimal technicalities, confining mathematical details to Section~\ref{sc:rigorous}.

We believe that the use of slip trajectories is a key conceptual step for developing a well-posed mathematical theory of dislocation motion, and this point of view seems to be different from all existing works we are aware of, including numerous classical approaches~\cite{Kondo55,Nye53,BilbyBulloughSmith55,Noll58,KronerSeeger59,Kroner60,Kroner61,Wang67}. In particular, it is distinct from the usual approach taken in the simulation methodology of Discrete Dislocation Dynamics~\cite{AmodeoGhoniem90a,AmodeoGhoniem90b,vanderGiessenNeedleman95,BulatovCai06book}, where is assumed that dislocation line segments can be assigned a pointwise velocity. There are several notable advantages of our approach: First, it allows to define the associated change in plastic distortion in a straightforward way. Second, there is a natural geometric notion which corresponds to the pointwise velocity of dislocation movement while containing more information. Third, working in a suitably general class of slip trajectories, we can evolve dislocations independently, even if they overlap or cross, which is not possible if considering \enquote{global} velocity fields (velocity fields \enquote{along the dislocation curves} require a parameterization, which is not readily available and also not preserved under the flow). As such, our approach provides a \enquote{weak formulation} of the dislocation motion, which we hope is sufficiently rich in structure to include all important physics and enable a fruitful rigorous analysis in the future (started in~\cite{Rindler21a?,Rindler21b?}).

Having outlined the description of dislocations in our model, we now turn to the connection between dislocation motion and the macroscopic mechanical properties of our specimen. Denoting the total deformation as $y \colon \Omega \subset \R^3 \to \R^3$, for which $\det \nabla y > 0$, the commonly used \emph{multiplicative} Kr\"{o}ner decomposition
\[
  \nabla y = E P
\]
splits the deformation gradient into elastic and plastic distortions $E, P$. Since $E, P$ are not in general deformation gradients themselves, they are henceforth referred to as \enquote{distortions}. We refer to~
\cite{Kroner60,LeeLiu67FSEP,Lee69EPDF,GreenNaghdi71,CaseyNaghdi80,GurtinFriedAnand10book,ReinaConti14,KupfermanMaor15,ReinaDjodomOrtizConti18,EpsteinKupfermanMaor20} for justifications and various other aspects of this decomposition. Our description of the crystal in Section~\ref{sc:scaffold} will in fact give its own justification of the Kr\"{o}ner decomposition based on the \emph{crystal scaffold}, which is the variable we use to describe the state of the crystal around a point. Namely, the crystal scaffold describes the bonds between the lattice atoms relative to the referential positions of the atoms. It is a purely kinematic quantity and can be defined unambiguously (unlike $P$, whose definition contains some ambiguities, which have to be explicitly excluded). We refer to Figure~\ref{fig:Schema} for a schematic overview of our model.

The elastic energy functional (assuming that our specimen is hyperelastic) only depends on $E$, i.e.,
\[
  \int_\Omega W_e(E) \dd x = \int_\Omega W_e(\nabla y P^{-1}) \dd x,
\]
where $W_e$ is the elastic energy density. Note that if $P$ is fixed and not a gradient (i.e., $\curl P \not\equiv 0$), the material may not be able to elastically relax to a stress-free configuration, see~\cite[Theorem~2.2]{LewickaPakzad11} and~\cite[Theorem~1.4]{KupfermanMaor15}. In this sense, our \enquote{reference configuration} should only be understood as a fixing of spatial coordinates and not as a configuration with special properties.

Our approach considers the plastic distortion $P$ to be an internal variable (like in~\cite{Mielke03a,MainikMielke05,FrancfortMielke06,MielkeMuller06,MainikMielke09}) and specifies that the evolution of $P$ occurs via the \emph{plastic flow equation}
\begin{equation} \label{eq:plastflow_intro}
  \dot{P} = D
\end{equation}
involving the \emph{total plastic drift} $D$, which acts as an infinitesimal (microscopic) generator of the flow of $P$. This in particular removes all ambiguity in the Kr\"{o}ner decomposition in our model, since we require that $P$ can change only through the motion of dislocations.

We propose that the plastic drift is expressed in terms of the movement of dislocations, i.e.\ via a slip trajectory. Namely, if a dislocation with Burgers vector $b$ moves in space-time along a slip trajectory $S$, then $D(t)$ is the $\R^{3 \times 3}$-valued measure (i.e., a \enquote{singular density}) on $\Omega$ given by
\begin{equation} \label{eq:D_from_disl}
  D(t) =  b \otimes g^b \, \Hcal^1 \restrict S^b|_t,
\end{equation}
where $S^b|_t$ is the slice of $S$ at time $t$, i.e.\ the dislocation line at time $t$, $\Hcal^1 \restrict S^b|_t$ is the (Hausdorff) measure concentrated on the dislocation line, and $g^b$ is the \emph{geometric slip rate}, which is given as
\[
  g^b := v^b\times\vec{T}^b,
\]
that is, the vector cross product of the \emph{dislocation velocity} $v^b \in \R^3$ and the dislocation line direction $\vec{T}^b$. Both $v^b$ and $\vec{T}^b$ can be obtained directly from $S^b$, but not from $S^b|_t$ alone (because the change in the time direction, i.e., the dislocation velocity, is missing). It therefore becomes clear that in order to define the plastic evolution we need to consider the complete \emph{path} of plastic evolution between two given points in time. Merely considering the endpoints of \enquote{elementary} movements as in the rate-independent models of~\cite{Mielke03a,MainikMielke05,FrancfortMielke06,MielkeMuller06,MainikMielke09} does not give us access to this information.

With the kinematics and dynamics specified, we then apply two fundamental principles of energetic modeling in mechanics, namely the \emph{Principle of Virtual Power}~\cite[Chapter~92]{GurtinFriedAnand10book} and the \emph{Free Energy Imbalance}~\cite[Section~27.3]{GurtinFriedAnand10book}, which is a consequence of the \emph{Second Law of Thermodynamics}. This yields formulas for the stresses corresponding to elastic and plastic distortions as well as the \emph{configurational stress} induced by dislocation motion. For the elastic and plastic stresses we recover (a version of) the classical Piola--Kirchhoff stress and the Mandel stress, respectively. For the configurational stress, that is, the stress power-conjugate to dislocation slip, we obtain an explicit expression, which is later seen to be a nonlinear analogue of the classical Peach--Koehler force.

The last ingredient of our model is a \emph{flow rule}, i.e.\ the constitutive relationship between rates and stresses, on the level of dislocations. A version of this approach is already employed in the Engineering literature on discrete dislocation dynamics, where \emph{mobilities} are typically prescribed~\cite{BulatovCai06book}. Here, we additionally connect this viewpoint with a fully nonlinear and frame-indifferent nonlinear elasticity theory. With a flow rule prescribed, our model is closed.

Let us remark that at some points this work makes some use of the language of currents (and with this, the language of differential geometry). In particular, this is the case when considering $2$-dimensional surfaces in $4$-dimensional space-time, where classical vector analysis has shortcomings (namely, there is no vector cross product in $\R^4$). While this is unavoidable for being precise in dealing with geometric objects, we have aimed to keep this to the absolute minimum necessary.

In this vein, one could have employed the so-called \enquote{geometrical language of continuum mechanics}~\cite{CiarletGratieMardare09,Epstein10book,EpsteinKupfermanMaor20} to obtain an additional level of consistency (and, perhaps, elegance), but this brings with it further notational complications. We have therefore opted not to pursue this avenue here. One noteworthy point in this respect is that the dislocation density of continuously distributed (and grain-homogenized) dislocations can be identified with the torsion tensor of an affine material connection~\cite{Kondo55,Nye53,BilbyBulloughSmith55,Noll58,KronerSeeger59,Kroner60,Kroner61,Wang67,Lazar00,Katanaev05,Malyshev07,HehlObukhov07,LazarAnastassiadis08,LazarAnastassiadis09,KupfermanMaor15,EpsteinKupfermanMaor20,KupfermanOlami20} (which can be seen as the limit of discrete dislocations~\cite{KupfermanMaor15,EpsteinKupfermanMaor20,KupfermanOlami20}). We refer to Appendix~\ref{ax:comparison} for a comparison of our approach to this geometric framework.

The present article does not aim to define all notions in a mathematically rigorous fashion and instead focuses on the physical modeling. However, we give some indication of how these notions can be made rigorous using the theory of integral currents in Section~\ref{sc:rigorous}. The actual mathematical implementation of the model is contained in two forthcoming works: First, the paper~\cite{Rindler21a?} develops a mathematical theory of space-time integral currents, which in our case represent the slip trajectories. In particular, the notion of \enquote{variation} (in time) is introduced and a number of results (e.g., on deformation, compactness, and comparison) are proved. Second, the work~\cite{Rindler21b?} proves an existence theorem for solutions to the rate-independent instance of our model, which can be interpreted as a validation of its mathematical consistency.

\subsection*{Outline}

In Section~\ref{sc:kinematics} we present the kinematic framework for plastic distortions and dislocations, before discussing their dynamics in Section~\ref{sc:dynamics}.  The energetic framework is described in Section~\ref{sc:energy}, where we also derive appropriate balance laws by applying fundamental thermodynamic principles. Section~\ref{sc:summary} summarizes all equations of the model and Section~\ref{sc:outlook} presents a discussion of various further effects that one may wish to incorporate, such as hardening. Finally, Section~\ref{sc:rigorous} is devoted to casting the model into a more rigorous setting based on the language of geometric measure theory. Appendix~\ref{ax:comparison} compares our approach to the classical \enquote{geometric paradigm} and \enquote{constitutive paradigm} of dislocation modeling.

\subsection*{Acknowledgements}

This project has received funding from the European Research Council (ERC) under the European Union's Horizon 2020 research and innovation programme, grant agreement No 757254 (SINGULARITY). Some parts of this work appeared in~\cite{Rindler15} in preprint form. The authors would like to thank Amit Acharya, Gilles Francfort, Cy Maor, Marco Morandotti, Michael Ortiz, and Florian Theil for helpful discussions related to the topic of this work. We have used the convention of ordering authors alphabetically, which is usual in the theoretical sciences.

\section{Kinematics} \label{sc:kinematics}

\begin{figure}
  \includegraphics[width=\linewidth]{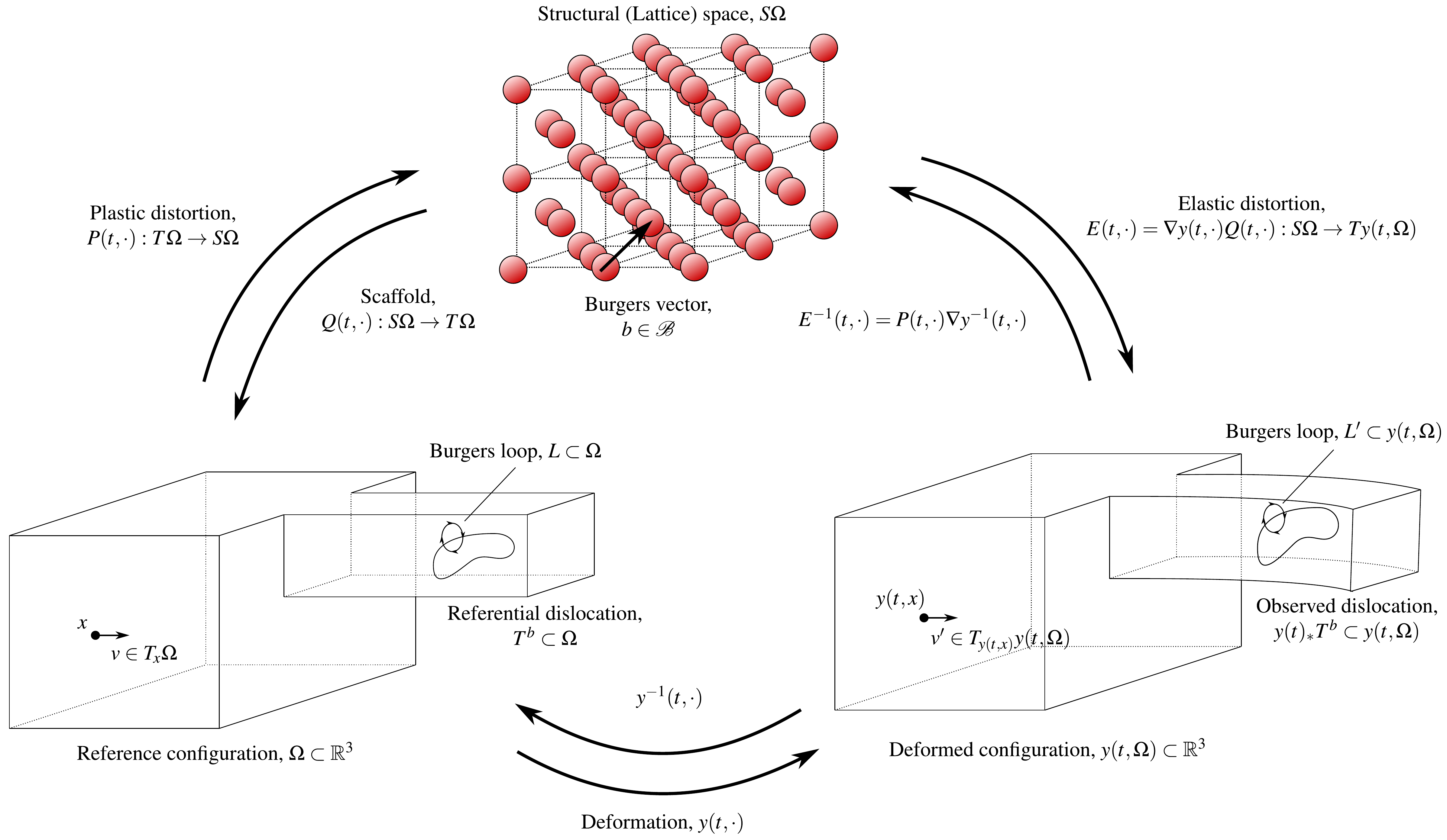}
  \caption{An overview of the kinematic framework.}
  \label{fig:Schema}
\end{figure}

We start by describing our approach to the modeling of the elastic, plastic, and dislocation kinematics of our specimen; a pictorial overview of our framework is provided in Figure~\ref{fig:Schema}.

\subsection{Basic modeling assumptions} \label{sc:crystal}

Consider a body formed of a single crystal, which will be allowed to be both elastically and plastically deformed. We state the fundamental principle of plasticity in crystalline materials as follows:
\begin{quotation}
  \textit{Plastic distortions manifest as changes to the crystal lattice, i.e.\ the breaking and reforming of bonds between the lattice atoms.}
\end{quotation}
Along with the usual description of material deformation through notions of strain, we will consider the local configuration of bonds in the crystal as a form of \emph{microstructure} attached to each material point. In particular, there are two important phenomena we keep track of:
\begin{enumerate}[(i)]
  \item \term{Slip}, i.e., the history of any rearrangement of interatomic bonds in the crystal.
  \item \term{Dislocations}, i.e., topological line defects in the crystal lattice, which cause a rearrangement of bonds to occur when they move through the material.
\end{enumerate}
These two features are described by \term{internal variables}, that is, additional quantities indexed by time $t$ and material points $x$. In particular, in agreement with experimental observation~\cite{HullBacon11book,AndersonHirthLothe17book}, we will assume within our modeling framework that these two phenomena are connected:
\begin{quotation}
  \textit{Slip occurs (only) through the motion of dislocations.}
\end{quotation}
We remark that focusing solely on these phenomena is valid as long as dislocations are relatively sparse and no other significant defects are present, such as a large number of point defects, grain boundaries, or cracks. Consideration of these lies beyond the remit of the present work.

\subsection{Deformation} \label{sc:macro}

We fix a Cartesian coordinate system on the 3-dimensional space within which the crystalline body is embedded; throughout this work, we will identify spatial points with their coordinate representation in $\R^3$ in this fixed coordinate frame. We then consider an open, bounded, connected \term{reference configuration} $\Omega \subset \R^3$ of material (or, more generally, a $3$-dimensional Riemannian body manifold $\Omega$). We denote the \term{referential (Lagrange, material) points} in $\Omega$ as $x = (x^1,x^2,x^3)$. Under a \term{(total) deformation} $y = (y^1,y^2,y^3) \colon \Omega \to \R^3$, every referential point $x$ is mapped to a \term{spatial (Euler) point} $y = y(t,x) \in y(t,\Omega)$, where the time $t$ is from an interval $[0,T]$. In the modeling that follows, we assume that $y$ and all other quantities are as smooth as required and we will frequently suppress function arguments for ease of reading.

A fundamental modeling assumption in this work (as in the continuum theory of elasticity and plasticity in general) is the following:
\begin{quotation}
  \textit{Macroscopically, the deformed body is again a continuum without interpenetration of matter or holes, and it has the same orientation as the reference configuration.}
\end{quotation}
In mathematical terms, this means that $y(t) = y(t,\frarg)$ is an orientation-preserving diffeomorphism, i.e., $y(t)$ is smooth, bijective, $\nabla y(t) \in \GL^+(3)$ in $\Omega$ (meaning $\det \nabla y(t) > 0$), and the inverse $y(t)^{-1}$ is itself smooth. Here, we denote the \term{(spatial) deformation gradient} as
\[
  \nabla y(t,x) = \left[ \frac{\partial y^j}{\partial x^k}(t,x) \right]^j_k \in \R^{3 \times 3}.
\]
We also remark at this point that derivatives with respect to the time variable $t$ of the various fields we consider will be denoted using the dot notation, i.e., $\dot{y} = \frac{\partial}{\partial t}y$.

A \term{referential vector} $v \in \Trm_x \Omega$ at a referential point $x \in \Omega$, where $\Trm_x \Omega \cong \R^3$ denotes the tangent space to $\Omega$ at $x \in \Omega$, is transformed into an \term{spatial (observed) vector} $w \in \Trm_{y(t,x)} y(t,\Omega)$ via the pushforward under $y$,
\[
 w := \nabla y(t,x) v.
\]
Physically, this is simply the statement that $\nabla y(t,x)$ encodes the net local transformation of the material close to a material point $x$ at some time $t$. While higher-order measures of strain could be considered, we will make the usual assumption (see, e.g.,~\cite{Ciarlet88}) that 
\begin{quotation}
  \textit{The deformation gradient is sufficient to describe the strain in the body we consider.}
\end{quotation}

\subsection{Crystal scaffold and plastic distortion}\label{sc:scaffold}

We now introduce two ways to describe the slip that has taken place up to a time $t$ at a referential point $x$, namely the plastic distortion $P$ and the crystal scaffold $Q$. Heuristically, the former describes the rearrangement of the atoms in the lattice due to slip, while the latter describes the rearrangement of the bonds. Both descriptions are mathematically equivalent, but from a modeling perspective they emphasize different aspects, and once we have described both approaches, we will freely transition between them.

\begin{figure}
  \centering
  \includegraphics[height=40mm]{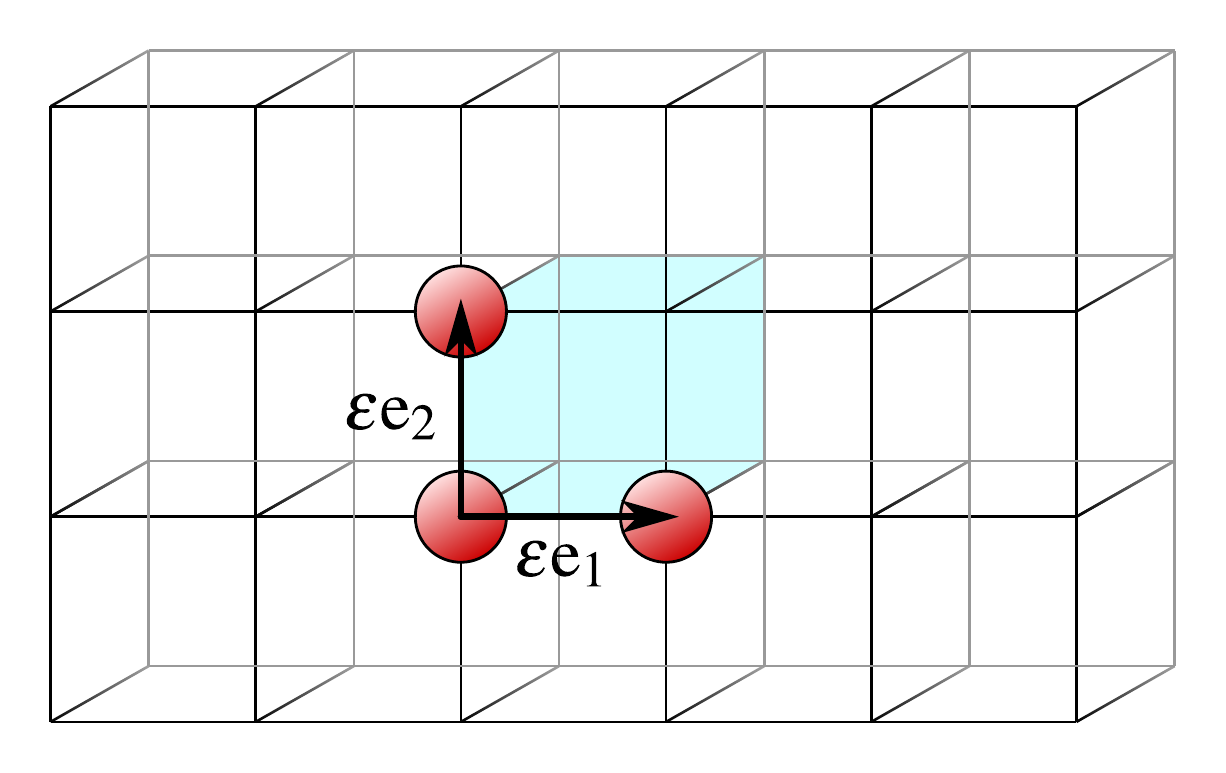}\\
  \includegraphics[height=40mm]{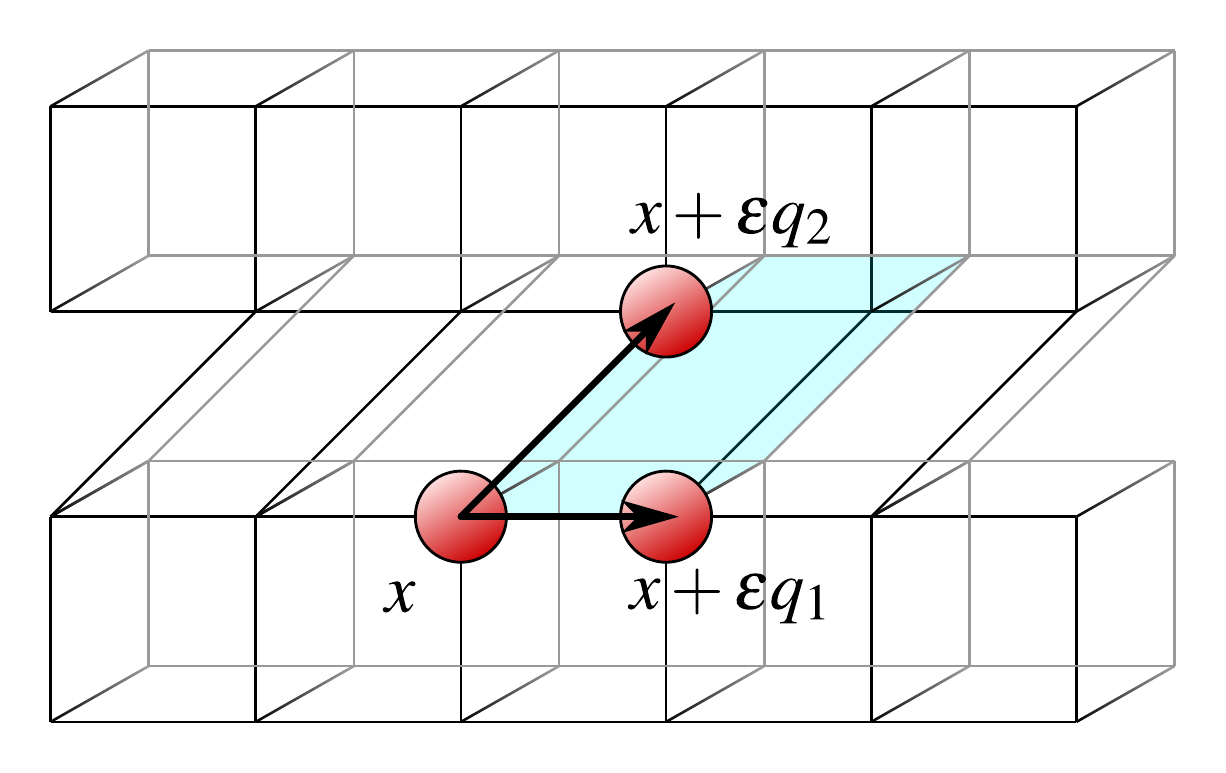}
  \includegraphics[height=40mm]{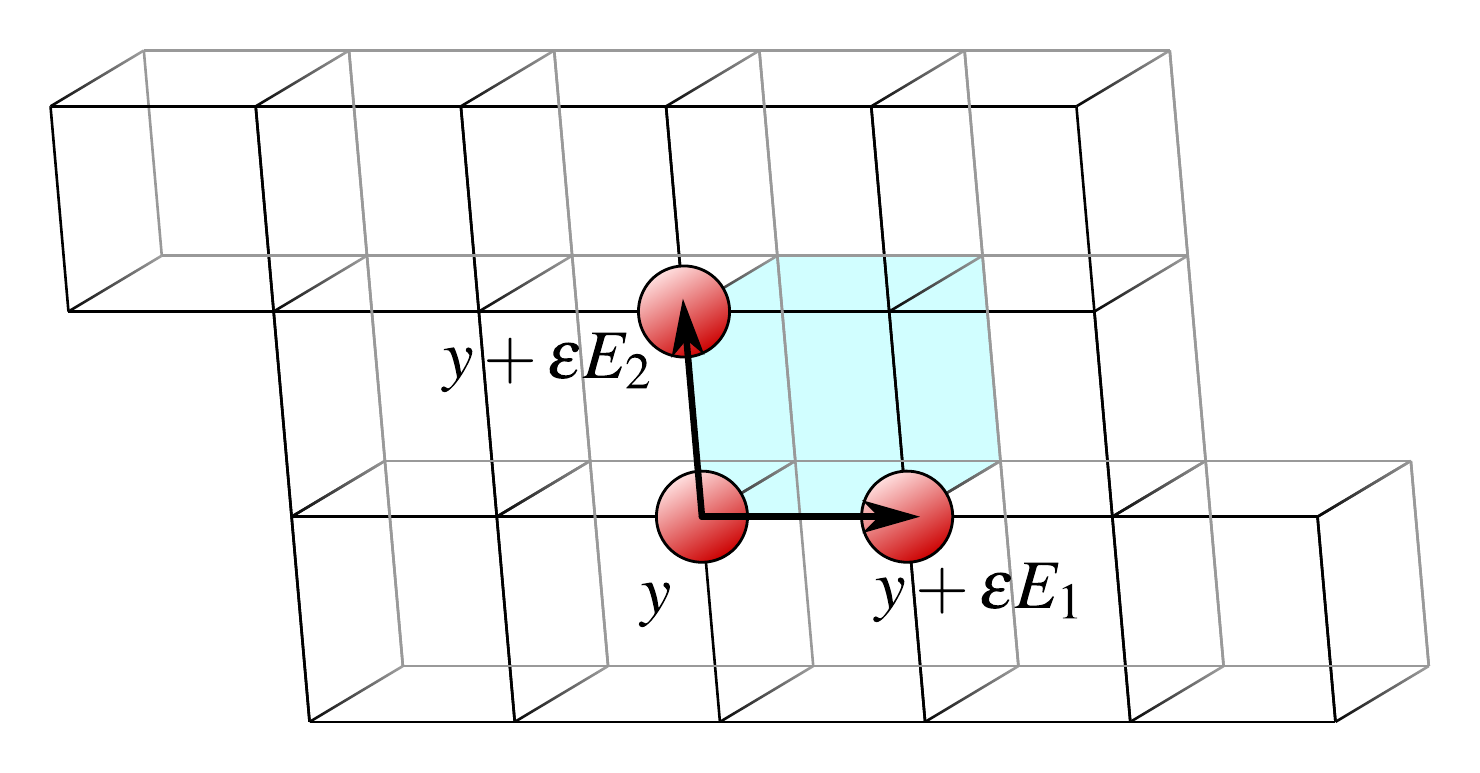}
  \caption{An illustration of scaffold vectors in a cubic lattice: The configuration corresponding to the bonds in the perfect lattice is shown at the top. The configuration after slip has occurred is shown on the lower left, where the atoms are held in place in their referential position and the scaffold vectors $q_i$ show the reconnected bonds. The further elastically deformed configuration is on the lower right.}
  \label{fig:scaffold}
\end{figure}

Let $\eps > 0$ be the microscopic lattice length scale (viewed as infinitesimal relative to our macroscopic length scale) and fix a material (Lagrangian) point $x$, which is mapped to a spatial (Euler) point $y$ by the deformation. For simplicity, suppose first that the underlying structure of the crystal is simply cubic with unit cell $[0,\eps]^3$. In a perfect crystal structure, the vectors $\eps \ee_1$, $\eps \ee_2$ and $\eps \ee_3$ therefore point to the nearest neighbor atoms in the lattice. Assume now that plastic slip has taken place, so that the crystal bonds have been rearranged, and the crystal cell has been elastically deformed. We denote the vector corresponding to the $i$th bond in the deformed configuration by $\eps E_i$ (these are the \enquote{arrows} connecting the atom at $y$ with the atoms that are bonded to it). Let us further assume that the atom that is now (after both slip and elastic deformation) at position $y + \eps E_i$ was previously at position $x + \eps q_i$ in the reference configuration. Since material cannot be infinitely stretched or compressed, we require that the vectors $\{E_i\}_i$ and $\{q_i|_x\}_i$ are bases with the same orientation as $\{\ee_i\}_i$, although these bases need not be orthogonal. Under the action of the deformation gradient, the vectors $q_i$ transform into the vectors $E_i$ (at least to leading order as the lattice scale $\eps$ vanishes). Thus, we may equate
\[
  \nabla y(x)q_i = E_i.
\]
We call the vectors $q_i$ the \term{scaffold vectors} of the crystal since they describe the bonds between the atoms (relative to the positions of the atoms in the reference configuration). We refer to Figure~\ref{fig:scaffold} for an illustration.

The vectors $q_i$ vary in time and with the referential point $x$ to which they are attached, and we collect them as the columns of the \term{crystal scaffold tensor (matrix)} $Q = Q(t,x) \in \GL^+(3)$. Similarly, we collect the vectors $E_i$ as the columns of the \term{elastic distortion tensor (matrix)} $E \in \GL^+(3)$. In this way,
\[
  \nabla y \,Q = E.
\]
We note that the above argument is, in fact, independent of the nature of the underlying lattice: for lattice types other than the simple cubic one, the vectors $q_i$ instead give the transformation of axes describing the unit cell.

It would be natural to assume that the $q_i$ are (possibly rotated) lattice vectors, so that $Q$ leaves the lattice invariant (forcing $Q$ to lie in the lattice point group). However, since we model our material as a continuum, we relax this assumption to allow more general scaffold matrices $Q\in\GL^+(3)$, assuming that the variation in the scaffold occurs on a length-scale below that of our continuum model. We discuss further restrictions on $Q$ in Section~\ref{sc:restrict} below.

Next, consider a mesoscopic (referential) piece $x + [0,\delta]^3$ of our specimen, where by \enquote{mesoscopic} we mean that $\delta > 0$ is small enough that the crystal scaffold vectors $q_i$ can be considered constant in $x$, but large enough that any defects in the lattice can be neglected. If this piece is cut out of the specimen, then the distorted lattice will seek to relax into its preferred shape, i.e. towards the undistorted lattice. This means that the crystal scaffold vector $q_i$ will be mapped back to $\ee_i$, and thus the piece will attain the shape $y + P([0,\delta]^3)$, where
\[
  P := Q^{-1}
\]
is the \term{plastic distortion tensor (matrix)}. While $Q$ describes the local bond structure between points in the reference configuration, $P$ describes how the points in the reference configuration may be locally embedded in the undistorted lattice structure of the material.  As $\nabla y \,Q = E$, this is equivalent to the multiplicative \term{Kr\"{o}ner decomposition}
\[
  \nabla y = E P,
\]
which is often considered as fundamental to all geometrically nonlinear modeling in finite-strain elasto-plasticity~\cite{Kroner60,LeeLiu67FSEP,Lee69EPDF,GreenNaghdi71,CaseyNaghdi80,GurtinFriedAnand10book,ReinaConti14}.

Note that in the argument above we have implicitly assumed that the crystal wants to restore the unit cell to its original shape, which is equivalent to the requirement that the (local) elastic energy density has a minimum only at the identity matrix $\Id$. If on the other hand this minimum should move to $\alpha \Id$ (for example through thermal expansion), then the local piece $x + [0,\delta]^3$ would instead take the shape $y + \tilde{P}([0,\delta]^3)$, where $\tilde{P} := \alpha Q^{-1}$. Moreover, if the identity matrix $\Id$ is a minimum of the energy density, then the whole point group $\Gfrak$ of the crystal will also be of minimal energy, and $PQ \in \Gfrak$ would also be an admissible relation. In contrast, $Q$ is a purely kinematic variable and makes no reference to an energy density, so may be considered more natural. In the following we will, however, avoid these ambiguities by \emph{defining} the plastic distortion tensor as $P := Q^{-1}$.

Moving back to the macroscopic point of view, in general
\[
  \curl P \neq 0,
\]
and indeed, this is a necessary property for the modeling of dislocations as topological lattice defects. Consequently, there may be no deformation $y^p$ with $\nabla y^p = P$, hence we only speak of $E$ and $P$ as the elastic and plastic \enquote{distortions}, respectively (this terminology is taken from~\cite{GurtinFriedAnand10book}). 

The model developed here will be based on an explicit flow for the crystal scaffold $Q$, or, equivalently, the plastic distortion $P$. Therefore, the \enquote{uniqueness problem} for $P$ (or $Q$), that is, the question which invariances should be required of $E, P$ in order to make the Kr\"{o}ner decomposition unique, which is much discussed in the literature~\cite{GreenNaghdi71,Rice71,Mandel73,NematNasser79,Dafalias87,LubardaLee81,Naghdi90,Zbib93,CaseyNaghdi80,Mielke03a}, is not relevant here.

\subsection{Burgers vectors}\label{sc:Burgersvec}

On a more formal mathematical level, the crystal scaffold $Q$ maps between two different types of vectors: \emph{structural vectors} and \emph{displacement vectors}. Let a vector $s$ be in \enquote{canonical lattice coordinates}, that is, \emph{assuming the crystal is perfect and uniform}, the vector $s$ represents a translation of $s^i$ units in the $i$th direction relative to some standard description of a particular crystal structure. As an example, in a cubic material, such a standard description might align the coordinate directions with the edges of a cubic cell. We term such vectors $s$ \term{structural vectors}, which lie in the \term{local structural space} $\Srm_x \Omega$ attached at $x$. Together, the structural spaces attached at all $x \in \Omega$ form the \term{(total) structural space} $\Srm \Omega := \bigcup_{x \in \Omega} \Srm_x \Omega$ (which has the mathematical structure of a vector bundle). Choosing to observe particular properties and symmetries of the structural vectors forms a modeling input into the choice of this space, since it reflects our belief about the ground-state lattice structure of the material.

On the other hand, points in the reference and deformed configurations are not taken from this perfect lattice, and are connected by \term{displacement vectors} in the ambient space within which the crystal sits. The spaces of displacement vectors are the tangent spaces $\Trm_x \Omega$ (in the reference configuration) and $\Trm_{y(x)} y(t,\Omega)$ (in the deformed configuration). We can use the matrix field $Q$ to map structural into referential displacement vectors and the matrix field $P$ for the reverse map. Likewise, $E = \nabla y\, Q$ maps structural vectors into spatial displacement vectors.

Since the fields $Q$ and $E$ provide local embeddings of the perfect lattice structure, we can detect certain topological defects in the structure of the lattice through them. In this context, defects are present when tracking structural vectors $s \in \Z^3$ over mesoscopic distances leads to non-unique referential or spatial offsets due to \emph{path-dependence}. Any discrepancy encountered is a \term{Burgers vector} as first introduced in dislocation theory in~\cite{Burgers39a,Burgers39b}.

\begin{figure}
  \includegraphics[width=\linewidth]{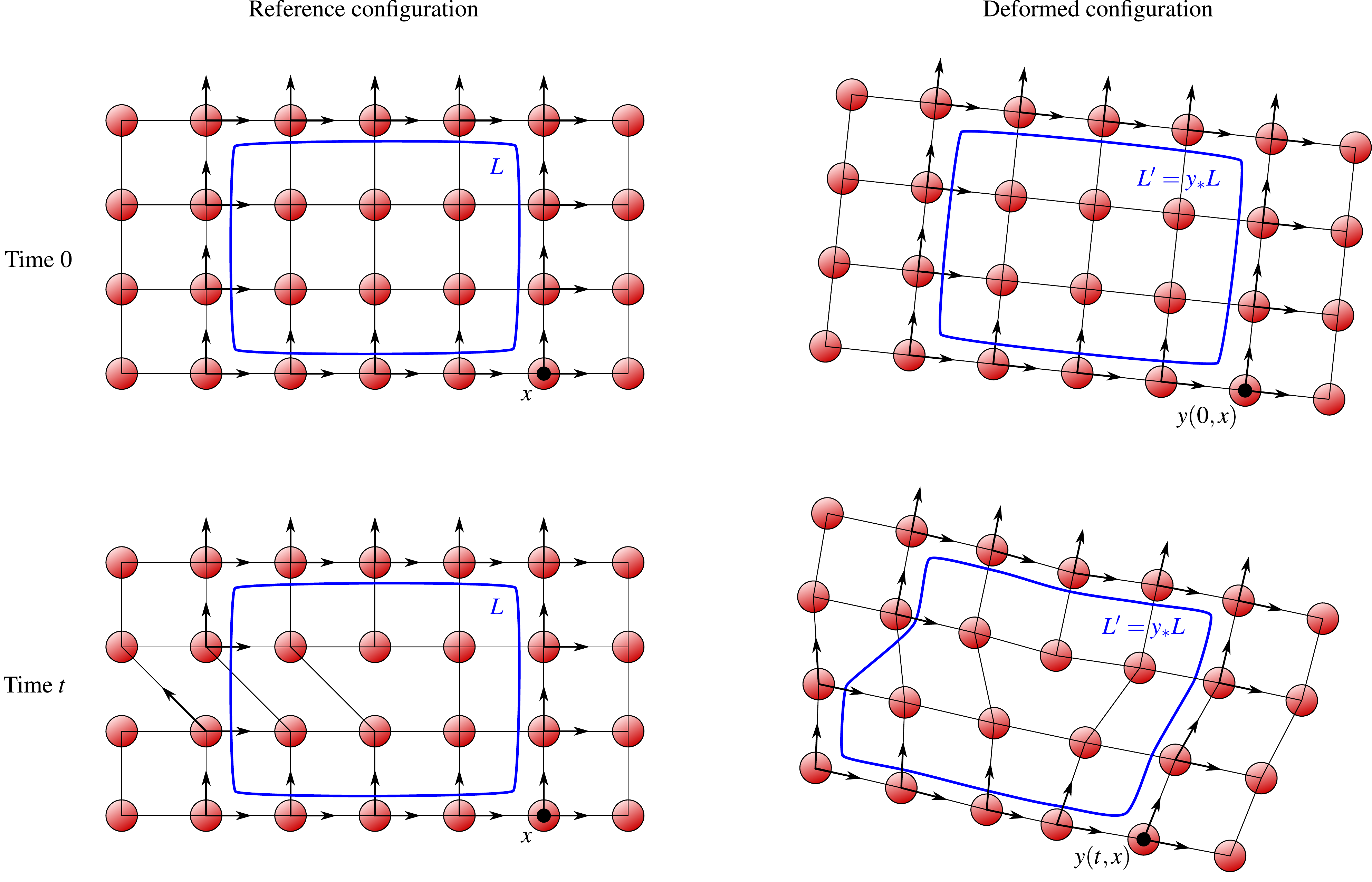}
  \caption{An illustration of loops around a dislocation used to find the Burgers vector.}
  \label{fig:Dislocation}
\end{figure}
To illustrate this idea further and make it more precise, consider Figure~\ref{fig:Dislocation}, which illustrates a dislocation lying in a plane within a simple cubic crystal. Choose two directions $\ee_1$ and $\ee_2$ in the reference configuration. At a referential point $x$ and time $t$, these map to lattice directions expressed by the structural vectors
\[
  p_1(t,x) := P(t,x)\ee_1, \qquad
  p_2(t,x) := P(t,x)\ee_2.
\]
Next, consider two paths in the reference configuration, starting at $x$. The first, $\ell_1$, involves traveling first $s^1$ units in the $\ee_1$-direction and then $s^2$ units in the $\ee_2$-direction. The second, $\ell_2$, involves traveling first $s^2$ units in the $\ee_2$-direction and then $s^1$ units in the $\ee_1$-direction. If $\tau$ is the unit tangent vector to these curves, then by integrating the structural vector density $P(t,x)\tau$ along each curve, we obtain a net structural vector which describes the total lattice displacement we experience relative to traveling through a perfect lattice. We define the discrepancy between the two results to be the Burgers vector around the referential loop $L := \ell_1-\ell_2$:
\[
  \text{Burgers vector around a referential loop }L = \int_{L} P\cdot \di s = \int_{\ell_1} P\cdot \di s-\int_{\ell_2} P\cdot \di s.
\]
If $S$ is the surface with normal $\ee_3$ enclosed by $L$, then, applying Stoke's theorem, we have that
\[
  \int_L P\cdot \di s = \int_S \curl P\cdot \di a,
\]
where $\curl = \nabla \times$ is taken row-wise. Letting the size of the loop tend to zero, we see that the Burgers vector density at $(t,x)$ over an infinitesimal surface with normal $\ee_3$ is $\curl P(t,x) \ee_3$. A more general construction shows that the Burgers vector density at $(t,x)$ over an infinitesimal surface with normal $N$ is $\curl P(t,x) N$. This corresponds precisely to Kr\"oner's dislocation density tensor $\alpha$~\cite{Kroner01,Kroner60,Kroner61}.

In the deformed configuration, we claim that the Burgers vector can be computed in a similar way, but using $E^{-1} = P (\nabla y)^{-1}$ in place of $P$. Since $y(t,\cdot)$ is assumed to be a diffeomorphism, denote the inverse of the deformation map $y(t,\cdot)$ as $x(t,\cdot)$. Then we assert that
\[
  \text{Burgers vector around a spatial loop }L' = \int_{L'} E^{-1}\big(t,x(t,y)\big)\cdot \di s'.
\]
We now demonstrate that this is equivalent to the referential definition above. Using the fact that $y$ is assumed to be a diffeomorphism, we have that any loop $L'\subset y(t,\Omega)$ is of the form $L' = y(t,L)$ for some loop $L\subset \Omega$, and so
\begin{align*}
  \int_{L'} E^{-1}(t,x(y))\cdot \di s'
  &=  \int_{L'} P(t,x(y))(\nabla y)^{-1}(t,x(y)) \cdot \di s'\\
  &= \int_{y(t,L)} P(t,x(y))\nabla x(t,y) \cdot \di s' \\
  &= \int_{L} P\cdot \di s.
\end{align*}
A similar argument as before entails that the Burgers vector density over an infinitesimal surface with normal $n$ located at the spatial point $y$ is $\curl E^{-1}(t,x(y))n$, where the curl is taken row-wise with respect to the spatial coordinates $y$. We note that a direct consequence of this derivation is that temporal changes in the deformation gradient $\nabla y$ alone do not affect the Burgers vector, whereas changes in $P$ (or, equivalently, $Q$) do.

We conclude this discussion by remarking that the engineering literature occasionally refers to the \enquote{structural (intermediate) configuration} between plastic and elastic distortions. This, however, can only be understood in the sense of \emph{vectors} and then only for locally perfect and uniform lattices, as explained above. It is meaningless to talk about \enquote{structural points} since the crystal scaffold $Q$ need not correspond to the gradient of an invertible map. Further, there is some disagreement in the literature over what should be called the \enquote{undistorted} crystal lattice, see for example~\cite[Section~91.2]{GurtinFriedAnand10book} and~\cite{Naghdi90}. The \emph{observed} lattice (which can be visualized for example through orientation-imaging microscopy) is the one that emerges after the new bonds have been formed and the atoms have relaxed their positions to an energetically optimal position. We here understand structural vectors relative to a locally perfect and uniform lattice, and so, the \emph{undistorted} lattice resides in the \enquote{structural space} (in agreement with~\cite[Section~91.2]{GurtinFriedAnand10book}).

\subsection{Incompressibility and restrictions} \label{sc:restrict}

We motivated the introduction of the scaffold tensor (matrix) $Q$ in Section~\ref{sc:scaffold} as a transformation which leaves the lattice invariant, but subsequently relaxed this in our continuum setting to allow for $Q \in \GL^+(3)$. Here, we demonstrate that by considering the lattice-based origins of $Q$, we can nevertheless deduce various restrictions we should require of this field.

If $\det Q \neq 1$ at some point in the crystal, then (in an averaged sense) the unit cell has grown ($\det Q > 1$) or shrunk ($\det Q < 1$) during slip. Physically, such motion requires dislocation climb to occur, which adds or removes a plane of lattice atoms~\cite[Chapter 3]{HullBacon11book}. This in turn requires interactions between dislocations and point defects, which we neglect in our present model. Thus, in our formulation, we postulate that \term{plastic incompressibility} holds, i.e.
\[
  \det Q = 1.
\]

In order to express further modeling restrictions, the plastic distortion $P$ or, equivalently, the crystal scaffold $Q$, may be restricted to a real matrix Lie group
\[
  \Pfrak \subset \SL(3) := \setb{ Q \in \R^{3 \times 3} }{ \det Q = 1 },
\]
that is, a topologically closed group of matrices (under matrix multiplication). We call $\Pfrak$ the \term{plastic distortion group}; see~\cite{Mielke02,Mielke03a,MainikMielke09} for earlier manifestations of this idea. So, we will require that
\[
  Q \in \Pfrak.
\]
If no further modeling restrictions beyond the incompressibility assumption are imposed, then $\Pfrak = \SL(3)$, corresponding to the plastic incompressibility above.

We denote the Lie algebra associated with the Lie group $\Pfrak$ by $\pfrak = \Lie(\Pfrak)$. It is defined to contain all those matrices $A \in \R^{3 \times 3}$ such that $\Exp(tA) \in \Pfrak$ for all $t \in \R$, where \enquote{$\Exp$} is the matrix exponential. More abstractly, $\Lie(\Pfrak)$ can equivalently be defined as the tangent space to $\Pfrak$ at the identity matrix, $\pfrak = \Trm_I \Pfrak$. Intuitively, an element of $\pfrak$ represents an \enquote{infinitesimal} plastic distortion, or, more precisely, a \term{plastic rate}, i.e.\ the speed by which a plastic distortion changes.

\begin{example}
The following are examples of physically-relevant Lie groups $\Pfrak$ and their corresponding Lie algebras $\pfrak$:
\begin{itemize}
  \item A single slip system with a unique slip direction $s \in \R^3$, $\abs{s} = 1$, and slip plane normal $n \in \R^3$, $\abs{n} = 1$ such that $s \perp n$, can be modeled using $\Pfrak = \set{ \Id + \alpha (s \otimes n) }{ \alpha \in \R }$ and its Lie algebra $\pfrak = \set{ \alpha (s \otimes n) }{ \alpha \in \R }$. This corresponds to the case where only one class of straight dislocations is permitted (the case considered in many dimension-reduced dislocation models, e.g., ~\cite{GarroniLeoniPonsiglione10,DLGP12,MSZ14,vMMP14,SPPG14,GvMPS16,CarrilloEtAl20}).
  \item In many materials which have a hexagonal close-packed structure, slip occurs only on planes with normal $n=(0,0,1)$, but in 6 different slip directions, which may be taken to be $(\pm1,0,0)$, $(\pm\tfrac12,\pm\tfrac{\sqrt{3}}2,0)$ and $(\mp\tfrac12,\pm\tfrac{\sqrt{3}}2,0)$. In this case, the slip directions have the same span as $(1,0,0)$ and $(-1,0,0)$, so we have
\[
  \Pfrak = \set{ \Id + \alpha \mathrm{e}_1 \otimes \mathrm{e}_3+\beta \mathrm{e}_2\otimes\mathrm{e}_3 }{ \alpha,\beta \in \R }.
\]
The relevant Lie algebra is $\pfrak = \set{ \alpha \mathrm{e}_1 \otimes \mathrm{e}_3+\beta \mathrm{e}_2\otimes \mathrm{e}_3 }{ \alpha,\beta \in \R }$.
  \item In the common high-symmetry cases of face-centred cubic (FCC) and body-centred cubic (BCC) materials, where dislocation glide is the only form of motion permitted, there are sufficiently many slip directions and normals to ensure that the appropriate Lie group is the maximal one, $\Pfrak = \SL(3) = \set{ A \in \R^{3 \times 3} }{ \det  A = 1 }$, which has the Lie algebra consisting of all deviatoric matrices, $\pfrak = \mathfrak{sl}(3) = \set{ A \in \R^{3 \times 3} }{ \tr\, A = 0 }$.
\end{itemize}
\end{example}

\subsection{Dislocations} \label{sc:disl}

As we have stated above, dislocations are line defects in the crystal lattice, which propagate slip through their motion. Their presence is characterized by a non-zero Burgers vector, which was discussed in Section~\ref{sc:Burgersvec} in the context of the plastic distortion. For a given material and physical regime, only a few possibilities for the Burgers vector of a dislocation are observed experimentally~\cite{HullBacon11book,AndersonHirthLothe17book}. To model this, we assume that there is a set of non-zero (structural) \term{Burgers vectors}
\[
  \Bcal = \bigl\{ \pm b_1, \ldots, \pm b_m \} \subset \R^3 \setminus \{0\}.
\]
Since structural vectors reflect the lattice structure of the material, the set $\Bcal$ should be assumed to be invariant under any point symmetries of the crystal (although this requirement is not important for the derivation to follow).

In the following we will often use the convenient notation
\[
  \int F(b) \dd \kappa(b) := \frac12 \sum_{b \in \Bcal} F(b)
\]
for any function $F$ defined on $\Bcal$. One may interpret $\kappa$ as the (purely atomic) \term{Burgers measure} $\kappa \in \Mcal^+(\R^3 \setminus \{0\})$ given as
\[
  \kappa := \frac12 \sum_{b \in \Bcal} \delta_b.
\]

A referential \term{dislocation system} is a collection
\[
  \Phi = (T^b)_{b\in\Bcal},
\]
where every $T^b$ is the set of dislocations with Burgers vector $b$ (which is constant along every slip trajectory, so both in space and time~\cite{HullBacon11book,AndersonHirthLothe17book}). So, $T^b$ consists of oriented curves in the reference configuration $\Omega$. Throughout this work, we will denote the consistently-oriented unit tangent vector to the collection of curves in $T^b$ as $\vec{T}^b$. More precisely, every $T^b$ can be expressed as a $1$-dimensional integral current in $\Omega$; see Section~\ref{sc:curr} for some elements of the theory of currents and Section~\ref{sc:geom} for the rigorous definition of dislocation systems.

As observed physically, we assume that the curves in $T^b$ are either closed loops or end at the boundary of the crystal, and switch orientation when the sign of the Burgers vector is switched. This can be expressed as
\begin{equation}\label{eq:Tb_props}
  \partial T^b = 0 \restrict \Omega, \qquad
  T^{-b} = -T^b 
\end{equation}
for all $b \in \Bcal$. Here, $\partial T^b$ is the boundary operator of $T^b$, which is the formal sum of the point masses at the end points of the curves in $T^b$ with their sign induced by the orientation of the respective curve ($+1$ at the \enquote{end} and $-1$ at the \enquote{start}). The symmetry condition entails that every dislocation occurs exactly twice in $\Phi = (T^b)_{b\in\Bcal}$, and hence we introduce a factor of $\frac12$ in the definition of $\kappa$ to cancel this double counting when integrating over $\kappa$. We could also consider dislocation systems which are not discrete, but we will not do so in the present work. 

If we want to transform to the deformed configuration, we need to consider the pushforward under the deformation, i.e., $y(t, T^b)$. Since we assume that $y(t,\frarg)$ is a diffeomorphism, it is easy to verify that $y(t,T^b)$ is boundaryless inside $y(t,\Omega)$, and so dislocations remain closed loops (inside $y(t,\Omega)$) in the deformed configuration.

\begin{figure}
  \centering
  \includegraphics[width=0.8\linewidth]{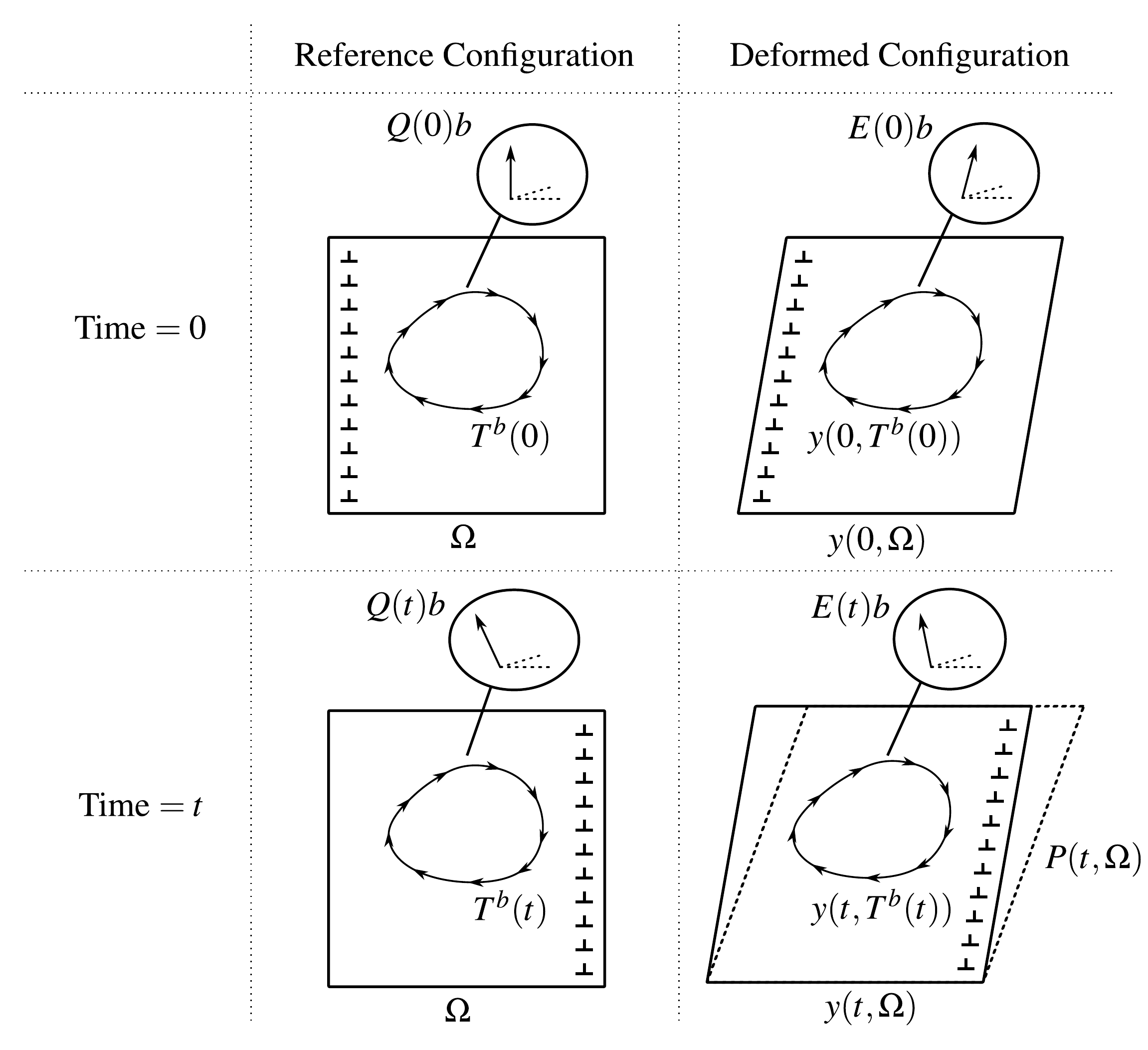}
  \caption{An illustration of the change to a dislocation loop under the movement of ambient edge dislocations.}
  \label{fig:BurgersVectorTransformation}
\end{figure}

We note that the character of the discrete dislocation loop $T^b$ with Burgers vector $b$ at a referential point $x$ is described by the relationship between $\vec{T}^b$ and $b$, but in our framework these vectors are of different type: $b$ is a structural vector but $\vec{T}^b$ is a referential displacement vector. The correct way to compare them is therefore to map $\vec{T}^b$ into the structural space via $P$, giving the curve tangent relative to the lattice. This leads us to say that if at a point $x$ the vectors $P\vec{T}^b$ and $b$ are orthogonal, then $T^b$ is an \emph{edge dislocation} at this point; if $P\vec{T}^b$ and $b$ are parallel then $T^b$ is an \emph{screw dislocation}; otherwise the dislocation is of \emph{mixed type}~\cite{HullBacon11book}. Note in particular that the movement of other ambient dislocations may change the type of a fixed dislocation loop since this movement changes $P$; see Figure~\ref{fig:BurgersVectorTransformation}.

Let us also remark that some works in the literature instead consider tensor-valued currents, see, e.g.,~\cite{ScalaVanGoethem19}. While this approach elegantly represents the symmetry $T^{-b}=-T^b$, it creates the issue that there is no uniqueness of decomposition into tensor products (at least in the case of \emph{fields} of dislocation lines, which we do not consider in the present work). Since our expression for the dissipation will also be allowed to depend on the Burgers vector $b$ (see Section~\ref{sc:energy} below), we choose the above $b$-indexed representation of dislocation systems instead.

\subsection{Thickened dislocation} \label{sc:thickened} 

When considering discrete dislocations as we do here, we are dealing with the case where the individual dislocations are \enquote{mascroscopically visible}. Thus, to accurately reflect their effect on the plastic distortion, we need to assign them a finite size. We do this by introducing a \term{dislocation line profile} $\eta \colon \R^3 \to [0,\infty)$ with compact support and satisfying $\int \eta \dd x = 1$. While a dependence of $\eta$ on $b$ or even on the tangent $\vec{T}^b$ to the line (allowing for different shapes of edge and screw dislocations) is conceivable, for simplicity, we consider $\eta$ to be fixed globally. Let \enquote{$\conv$} denote convolution in the spatial variables. Then, the \term{thickened dislocation system} is defined as $\Phi_\eta := (T^b_\eta)_b$ with
\[
  T^b_\eta := \eta \conv T^b,
\]
that is, $T^b_\eta$ is \enquote{smeared out} with shape $\eta$ (in space). The approach of assigning dislocations a finite size within continuum models is common; see for example~\cite{CaiArsenlisWeinbergerBulatov06,ContiGarroniOrtiz15}.

\section{Dynamics}
\label{sc:dynamics}

Now that we have established a description of the kinematic quantities, we next present our approach to a dynamical theory.

\subsection{Slip trajectories and velocity} \label{sc:velocity} 
In order to make clear mathematical sense of dislocation movement within our framework, we will consider the evolution of the loops $T^b$ in (Galilean) space-time $\R^{1+3} \cong \R \times \R^3$, where the first component takes the role of \enquote{time} and the remaining components take the role of \enquote{space}. The unit vectors in $\R^{1+3}$ are denoted by $\ee_0, \ee_1, \ee_2, \ee_3$ with $\ee_0$ the \enquote{time} unit vector (pointing in positive direction). We also define orthogonal projections onto the time and space components as
\[
  \tbf(t,x_1,x_2,x_3) := t \qquad\text{and}\qquad
  \pbf(t,x_1,x_2,x_3) := (x_1,x_2,x_3),
\]
respectively. On occasion, we will consider spatial vectors in $\R^3$ as space-time vectors in $\R^{1+3}$ by extending them by zero in the $\ee_0$ direction, i.e., identifying 
\[
  (x_1,x_2,x_3)\in\R^3\quad\text{with}\quad (0,x_1,x_2,x_3)\in\R^{1+3}.
\]

\begin{figure}
  \includegraphics[width=0.6\linewidth]{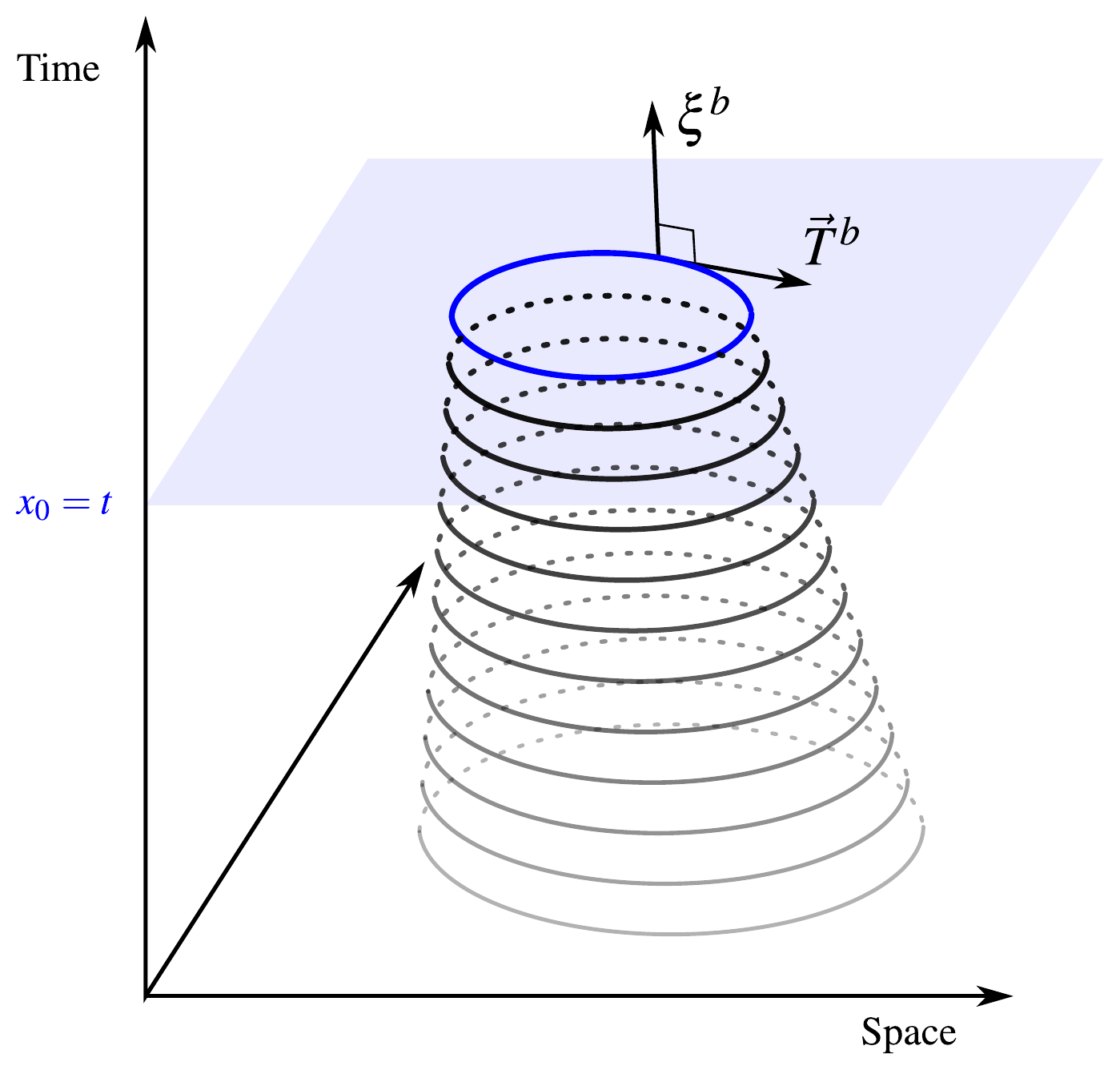}
  \caption{An illustration of a slip trajectory showing the evolution of a single loop in time, along with accompanying tangent vectors (for visualization purposes only two spatial dimensions are shown). }
  \label{fig:SlipTrajectory}
\end{figure}

A referential \term{slip trajectory system} over the time interval $[0,T]$ is a collection
\[
  \Sigma = (S^b)_{b\in\Bcal},
\]
where each $S^b$ is a set of $2$-dimensional oriented surfaces lying in the $4$-dimensional space-time cylinder $[0,T]\times \Omega$. In this way, $S^b$ collects the trajectories of all dislocations with Burgers vector $b$. A precise mathematical formulation models the $S^b$ as $2$-dimensional integral currents in space-time, which enables one to put the following derivation on a rigorous footing; see Section~\ref{sc:rigorous} for further details.  An illustration of slip trajectories is given in Figure~\ref{fig:SlipTrajectory}.

We also assume
\begin{equation} \label{eq:Sb_sym}
  S^{-b} = -S^b \qquad\text{for every }b\in\Bcal
\end{equation}
and that $S^b$ has no boundary on the interior of the space-time cylinder, i.e.
\begin{equation} \label{eq:Sb_bdry}
  \partial S^b \restrict \big( (0,T) \times \Omega\big) = 0,
\end{equation}
where the symbol \enquote{$\restrict$} denotes the restriction. This assumption is a natural extension of the requirement that dislocations are composed of loops inside $\Omega$, and that these loops are not instantaneously created or destroyed, but must instead grow from or shrink to a point. As a consequence of these assumptions, for all $t\in[0,T]$, restricting $S^b$ to the time-slice $\{t\}\times\Omega$ (and projecting onto $\R^3$) results in the dislocations described by $T^b(t)$ satisfying the properties assumed in Section~\ref{sc:disl}. More formally, one obtains $T^b(t)$ from $S^b$ by considering the slice $S^b|_t$ (with respect to the time projection $\tbf$) and pushing forward under $\pbf$, giving the definition $T^b(t) := \pbf_*(S^b|_t)$.

Since slip trajectories are space-time surfaces, they have (two-dimensional) space-time tangent spaces. As we will see, the geometric properties of the tangent space to $S^b$ are directly connected to properties of the dislocations whose motion they represent. Indeed, the (spatial) tangent vector $\vec{T}^b(t,x)$ to $T^b(t)$ at $x\in\Omega$, can be lifted to the (space-time) tangent space of $S^b$ at $(t,x)$:
\[
  \vec{T}^b(t,x)\in \Trm_{(t,x)} S^b.
\]
Moreover, since $|\vec{T}^b|^2=1$, we see that the (space-time) vector $\vec{T}^b$ can be used as the first element of an orthonormal basis for the tangent space $\Trm_{(t,x)}S^b$. We denote the second orthonormal basis vector by $\xi^b(t,x)\in \Trm_{(t,x)}S^b$, which is uniquely defined once we require that it points \enquote{forward in time}, i.e., that
\[
  \tbf(\xi^b(t,x)) = \xi^b(t,x)\cdot \ee_0\geq 0.
\]
Hence,
\[
  \Trm_{(t,x)}S^b = \spn\big\{\vec{T}^b(t,x),\xi^b(t,x)\big\}.
\]
Physically, the above statement says that the tangent space to a slip trajectory $S^b$ at a point $(t,x)$ is the span of the tangent vectors to the dislocations passing through $x$ at time $t$, which (by definition) have no components in the time direction, along with $\xi^b$, which contains all information about the rate and direction of motion.

We also define the \term{(referential) dislocation velocity} of $S^b$ at $(t,x)$ as
\[
  \frac{\DD}{\DD t} S^b(t,x) := v^b(t,x) := \frac{\pbf(\xi^b(t,x))}{\abs{\tbf(\xi^b(t,x))}} \in  \R^3.
\]
To ensure that this is well-defined, we in fact require that all slip trajectories satisfy the following regularity condition:
\[
  \tbf(\xi^b(t,x)) = \xi^b(t,x)\cdot\ee_0 > 0.
\]
This means that the slip surface has no \enquote{vertical} parts. With $\pbf(\xi^b)$ being the spatial displacement of the dislocation (locally around a point) per $|\tbf(\xi^b)|= \xi^b \cdot \ee_0$ units of time, the spatial velocity of the dislocation is indeed given by the above definition of $\frac{\DD}{\DD t} S$. We further note that the dislocation velocity $v^b$ is orthogonal to the tangent vector $\vec{T}^b$ since, by construction, $\xi^b$ and $\vec{T}^b$ are orthogonal, and hence
\[
    0 = \xi^b \cdot \vec{T}^b = \pbf(\xi^b)\cdot \vec{T}^b = \tbf(\xi^b) \, v^b\cdot \vec{T}^b,
\]
where we have considered $\vec{T}^b$ both as a vector in $\R^3$ and $\R^{1+3}$. This fact expresses that the dislocation velocity does not have a component along the tangent of the curve (which would have no meaning). We further refer to Section~\ref{sc:slipvel} for a more geometric view on the definition of the dislocation velocity.

\subsection{Dislocation motion and infinitesimal plastic shear} 

A central ingredient of our model is an expression for the change in the crystal scaffold caused by a dislocation traveling along a slip trajectory. Concretely, a dislocation with Burgers vector $b\in\Bcal$ moving within a structural plane $H'$ in the perfect lattice causes a rearrangement of bonds as follows: A bond $s$ which has a component in the normal direction to $H'$ is broken and reconnected to an adjacent lattice point $s-b$ (this choice of sign seems to be the most natural one, as will be clear in the following). This action corresponds to applying a structural shear of the form $\Id - b\otimes N'$ to each bond where $N'$ is the structural normal to the plane $H'$. The action of this rearrangement of bonds corresponds to mapping the referential scaffold vector $q = Qs$ to $q-Qb$. We thus posit that the motion of a single dislocation across the plane $H'$ yields a net transformation of scaffold vectors via
\[
  q\mapsto \bigl(\Id - (Qb)\otimes N\bigr)q,
\]
where $N = Q^{-T}N'$ now is the referential normal vector corresponding to the structural normal vector $N'$. In this context note that we need to transform $N'$ to $N$ taking into account the dilation of the scaffold vectors, so that $N = Q^{-T}N'$ need not be of unit length: the structural shear $\Id - b\otimes N'$ applied to $Q^{-1}q$ yields precisely the above expression.

If the dislocation velocity is $\frac{\DD}{\DD t} S^b = v^b$ then the normal vector $N$ can be expressed as
\begin{equation}\label{eq:slipnorm}
  N = \frac{v^b\times \vec{T}^b}{|v^b\times \vec{T}^b|} = \frac{v^b\times \vec{T}^b}{|v^b|},
\end{equation}
where the second equality follows from the standard result that $|a\times b| = |a| |b|$ for orthogonal vectors $a,b\in\R^3$.
More generally, if $\phi^b$ dislocations with Burgers vector $b$ move within the plane $H$, the crystal scaffold $Q$ is transformed via the \term{infinitesimal plastic shear} relation
\begin{equation} \label{eq:DtR}
  Q \mapsto Q' := \bigl( \Id - (Qb) \otimes N \, \phi^b \bigr) Q   \qquad\text{at time $t$ such that $(t,x) \in \supp T^b(t)$.}
\end{equation}
We note that~\eqref{eq:DtR} implicitly assumes that lattice planes transform in a straightforward way under the scaffold map $Q$.  This, in fact, turns out to be an issue which is slightly more subtle than it might appear at first sight. A full derivation is presented in Section~\ref{sc:slipderivation}.

\subsection{Plastic flow equation} \label{sc:plastflow}
 
We will now derive the plastic flow equation from the infinitesimal plastic shear relation~\eqref{eq:DtR}. We refer to Section~\ref{sc:plastflow_rigorous} for more geometrically rigorous arguments.

Recall the definition of the thickened dislocation system $(T^b_\eta)_b$ from Section~\ref{sc:thickened} and also define the \term{thickened slip trajectories} $\Sigma_\eta := (S^b_\eta)_b$ as
\[
  S^b_\eta := \eta \conv S^b.
\]
Consider an infinitesimal time interval $[t,t+\delta]$ and a referential point $x\in\Omega$. Over the course of this time interval, the amount of dislocations with Burgers vector $b$ passing through (near) this point is
\begin{equation}\label{eq:PartDensity}
    \phi^b(x;[t,t+\delta]) = \int_t^{t+\delta} \, m^b_\eta(\tau,x) \, |v^b_\eta(\tau,x)| \dd \tau,
\end{equation}
where $m^b_\eta(s,\frarg)$ is the \term{multiplicity} of $T^b_\eta(s)$, that is, the (vector) norm of the density of $T^b_\eta(s)$ (with respect to Lebesgue measure), and $|v^b_\eta|$ is the speed at which the dislocations are transported past the point $x$. The expression in~\eqref{eq:Qtks} for $\phi^b$ is equal to the value $\abs{\pbf(S^b_\eta)}([t,t+\delta] \times \Omega)$, where we wrote the \enquote{total slip measure} as $\abs{\pbf(S^b_\eta)} := \abs{\pbf(\vec{S}^b_\eta)} \, \tv{S^b_\eta}$; this follows from the rigorous arguments made in Section~\ref{sc:plastflow_rigorous}.
  
Combining the infinitesimal plastic shear relation~\eqref{eq:DtR} (for the thickened slip trajectories) with the expression for $\phi^b$ in~\eqref{eq:PartDensity}, it follows that
  \begin{equation} \label{eq:Qtks}
    Q(t+\delta) \approx \biggl(\Id -  (Q(t) b) \otimes N_\eta \int_t^{t+\delta} m^b_\eta(\tau) \, |v^b_\eta(\tau)| \dd \tau \biggr) Q(t).
  \end{equation}
Rearranging,
\[
  - Q(t)^{-1} \frac{Q(t+\delta) - Q(t)}{\delta} Q(t)^{-1} \approx b \otimes \frac{N_\eta}{\delta} \int_t^{t+\delta} m^b_\eta(\tau) \, |v^b_\eta(\tau)| \dd \tau,
\]
so we may take the limit as $\delta\to0$ to arrive at
  \begin{equation}
  -Q^{-1}\dot{Q}Q^{-1} = b \otimes g^b, \label{eq:QQinv}
\end{equation}
where we defined the \term{geometric slip rate} to be
\begin{equation}\label{eq:gb}
  g^b := N_\eta \, m^b_\eta \, |v^b_\eta|
  = v^b_\eta \times \vec{T}^b_\eta \; m^b_\eta
  = \eta \conv \big[ v^b \times \vec{T}^b \, m^b\big]
  \in \R^3,
\end{equation}
with $m^b$ the multiplicity of dislocations lines in $T^b$ (which really is a singular measure with respect to Lebesgue measure) at a point. Here we used~\eqref{eq:slipnorm} for the first equality. The second equality in~\eqref{eq:gb} follows because the mollification commutes with the operation of computing the geometric slip rate (see Section~\ref{sc:plastflow_rigorous}). Intuitively, the geometric slip rate $g^b$ is the normal vector field to the dislocation motion, with magnitude proportional to the velocity and number of flowing dislocations.

Finally, several slip systems with different Burgers vectors $b$ may be active at the same time. We assume that these are additive at the level of plastic flow, and thus we obtain the \term{plastic flow equation}
\begin{equation} \label{eq:plast_flow}
  \dot{P} = -Q^{-1}\dot{Q}Q^{-1} = D
\end{equation}
with the \term{total plastic drift} 
\begin{equation} \label{eq:D}
  D(t,x) := \int b \otimes  g^b(t,x) \dd \kappa(b) \in \R^{3 \times 3}.
\end{equation}

Alternatively, and this is the approach taken in Section~\ref{sc:plastflow_rigorous}, one may first define the $2$-vector-valued geometric slip rate
\begin{equation}\label{eq:gammab}
  \gamma^b(t,x) := \eta \conv \biggl[ \frac{\DD}{\DD t} S^b(t,\frarg) \wedge \vec{T}^b(t) \; m^b \biggr],
\end{equation}
which arises naturally as the quantity describing the flow of dislocations via the coarea formula. We note that a $2$-vector $a\wedge b$ represents the oriented plane spanned by the vectors $a$ and $b$, which in $3$-dimensional space (but not in $4$-dimensional space-time), can equivalently be represented by its oriented normal $a \times b$. In $\R^3$, the transformation between $2$-vectors and normal vectors is given via Hodge duality, and $g^b$ defined in~\eqref{eq:gb} is indeed the Hodge dual of $\gamma^b$.

\begin{example} \label{ex:SL}
Assume just plastic incompressibility, that is, $\Pfrak = \SL(3)$, so $\det Q = 1$. Then, $\pfrak = \mathfrak{sl}(3)$, the vector space of deviatoric (i.e., trace-free) matrices. Using Jacobi's formula and Cramer's rule (whereby $Q^{-1} = (\cof Q)^T$) as well as~\eqref{eq:plast_flow}, we compute
\[
  \frac{\di}{\di t} \det Q
  = \cof Q : \dot{Q}
  = \tr(Q^{-1} \dot{Q})
  = -\tr (DQ)
  = -\tr (QD).
\]
Thus, if the initial value $Q_0$ for $Q$ satisfies $\det Q_0 = 1$ (everywhere in $\Omega$), then for the preservation of this property, $QD$ needs to be deviatoric, i.e.
\[
  \tr (QD) = 0.
\]
Using the formula~\eqref{eq:D} for $D$, this holds if $Qb$ is orthogonal to $g^b$ (the normal to the slip trajectories), which is the case precisely if the dislocation motion is a \emph{glide}~\cite{HullBacon11book,AndersonHirthLothe17book} since $Qb$ is the referential manifestation of the (structural) Burgers vector $b$. If there is dislocation \emph{climb}, then $(Qb) \cdot g^b \neq 0$ and the plastic flow is not volume-preserving. We note that it is natural that this condition depends on $Qb$ and not $b$ directly: In a distorted crystal the scaffold has changed and so the referential slip planes along which the crystal can deform in a volume-preserving fashion, have changed with it; for an illustration, see Figure~\ref{fig:BurgersVectorTransformation}. We refer to Section~\ref{sc:climb} for a brief discussion of how climb might be incorporated into the present model.
\end{example}

\subsection{Consistency} \label{sc:consist}

In our model, there are two different ways of detecting dislocations: First, the dislocations are given explicitly in the dislocation system $(T^b(t))_b$ at time $t$. Second, they can be detected by computing the circulation of $P$ around the dislocation and adding up tangent (lattice) vectors, as explained in Section~\ref{sc:Burgersvec}. This is equivalent to $P$ having a non-trivial curl, i.e.
\[
  \curl P \neq 0.
\]
For the consistency of our model it is important to verify that the defects expressed in $P(t)$ remain the same as the ones expressed in the dislocation system $(T^b(t))_b$. Indeed, the dynamics can be shown to preserve the \term{consistency condition}
\begin{equation} \label{eq:consist}
  \curl P(t) = \int  b \otimes T^b_\eta(t) \dd \kappa(b)  \qquad\text{in $\Omega$}
\end{equation}
along the flow. This means that~\eqref{eq:consist} holds for all $t \in (0,T)$ if it is satisfied at $t = 0$, where we assume it as a condition on the initial data. A full proof of~\eqref{eq:consist} requires the use of geometric tools, and so the details are postponed to Section~\ref{sc:consistproof}.

\section{Energetics} \label{sc:energy}

We next apply two fundamental principles of energetic modeling in mechanics. First, for the \emph{Principle of Virtual Power}~\cite[Chapter~92]{GurtinFriedAnand10book} one considers all \enquote{virtual} motions of the system under consideration, that is, all allowed motions within the independent degrees of freedom. These motions are referred to as \enquote{virtual} because they might not be attained in a given evolution (but they are \emph{attainable}). Concretely, the motions our system can undergo are the variations in the total deformation $y$ of the specimen and the movement of the dislocations via a slip trajectory $\Sigma$. The other quantities $P$ (or $Q$) and $E$ only change as a consequence of movement in $y$ and the slip trajectory $\Sigma$. Second, we will invoke the \emph{Free Energy Imbalance}~\cite[Section~27.3]{GurtinFriedAnand10book}, which is a version of the \emph{Second Law of Thermodynamics}, to obtain a relation on the directionality of the evolution.

\subsection{Energy and stresses}

For a deformation $y$, which is an orientation-preserving diffeomorphism and hence satisfies $\det \nabla y > 0$ in $\Omega$, and a scaffold map $Q$, we define the \term{elastic energy} via
\begin{equation} \label{eq:W}
  \Wcal_e(y,Q) := \int_\Omega W_e(\nabla y Q) \dd x,
\end{equation}
where $W_e \colon \GL^+(3) \to \R$ is the \term{elastic energy density}. This shape of the energy functional is explained as follows: Postulating the Cauchy--Born rule~\cite{Ericksen08,EMing07a,EMing07b,HudsonOrtner12,OrtnerTheil13}, the elastic potential energy of the deformation will depend on $E = \nabla yQ$ only, since $E$ encodes the stretching of the bonds in a crystal unit cell.

We further require $W_e$ to satisfy the \term{objectivity (frame-indifference)} condition
\begin{equation} \label{eq:frame_indiff}
  W_e(RE) = W_e(E)  \qquad\text{for all $R \in \SO(3)$, $E \in \GL^+(3)$.}
\end{equation}
Often, $W_e$ also reflects additional material symmetries, such as $W_e(EZ) = W_e(E)$ for all $Z$ from a \term{(point) symmetry group} $\Gfrak \subset \SL(3)$ of the crystal. A common material symmetry is \term{isotropy}, for which $\Gfrak = \SO(3)$, see, e.g., ~\cite[Section~3.4]{Ciarlet88} and~\cite{GrandiStefanelli17}. Note that for $Z \in \Gfrak$ we have $EZ = \nabla y QZ$ and the corresponding scaffold $Q' := QZ$ represents a symmetry transformation of the lattice.

Let $f(t) = f(t,x) \in \R^3$ be an external \term{bulk loading}; here, we avoid explicitly considering other types of loading to minimize complexity, but there is of course no significant restriction on the form this can take. We then define the \term{total energy} as
\begin{equation} \label{eq:E}
  \Ecal(t,y,Q) := \Wcal_e(y,Q) - \int f(t) \cdot y \dd x.  
\end{equation}

Let $\Omega' \subset \Omega$ be a referential subdomain of the body $\Omega$. The \term{internal power} expended within $\Omega'$ is given as
\[
  \Ical(\Omega') = \int_{\Omega'} \frac{\di}{\di t} W_e(\nabla y Q) \dd x + \int \int_{\Omega'} X^b\cdot g^b \dd x \dd \kappa(b)
\]
where $X^b \in \R^3$ ($b \in \Bcal$) is the \term{referential configurational stress} that is power-conjugate to the geometric slip rate $g^b$. Section~\ref{sc:PKstress} below will show how $X^b$ can be considered a nonlinear analogue of the Peach--Koehler force~\cite{PeachKoehler50}.

Assuming that we may differentiate freely, we compute
\[
  \dot{E} = \nabla\dot{y}Q + \nabla y \dot{Q}
  = \nabla\dot{y}Q - \nabla y Q L,
\]
where
\[
  L := \dot{P}P^{-1} = -Q^{-1} \dot{Q}
\]
is the \term{structural plastic rate}. Then,
\begin{align*}
  \frac{\di}{\di t} W_e(\nabla y Q)
  &= \DD W_e(E) : \dot{E}\\
  &= \DD W_e(\nabla y Q) Q^T : \nabla \dot{y} - Q^T \nabla y^T \DD W_e(\nabla y Q) : L \\
  &= T : \nabla \dot{y} - M : L,
\end{align*}  
where we have defined the \term{Piola--Kirchhoff stress} (referential elastic stress) as
\[
  T := \DD W_e(\nabla y Q) Q^T
\]
and the \term{Mandel stress} (structural plastic stress) as
\begin{equation} \label{eq:Mandel}
  M := Q^T \nabla y^T \DD W_e(\nabla y Q) = E^T \DD W_e(E).
\end{equation}
We observe that the structural plastic drift $L$ can be related, via the plastic flow equation~\eqref{eq:plast_flow}, to the referential plastic drift given in~\eqref{eq:D} as follows: 
\[
  L = -Q^{-1} \dot{Q} Q^{-1} Q
    = D Q
    = \biggl( \int b \otimes g^b \dd \kappa(b) \biggr) \; Q,
\]
and thus
\begin{align*}
  \int_{\Omega'} M : L \dd x
  &= \int \int_{\Omega'} [M Q^T] : [b \otimes g^b]\dd x \dd \kappa(b) \\
  &= \int \int_{\Omega'} \big(Q M^T b\big)\cdot g^b \dd x \dd \kappa(b).
\end{align*}
Combining the above considerations and also using the Gauss--Green theorem,
\begin{align*}
  \Ical(\Omega') &= \int_{\Omega'} T : \nabla \dot{y} - M : L \dd x + \int \int_{\Omega'} X^b\cdot g^b \dd x \dd \kappa(b)\\
  &= - \int_{\Omega'} \Diverg T \cdot \dot{y} \dd x + \int_{\partial \Omega'} T n \cdot \dot{y} \dd a - \int \int_{\Omega'}  \big(QM^Tb\big)\cdot g^b \dd x \dd \kappa(b) \\
  &\qquad + \int \int_{\Omega'} X^b\cdot g^b \dd x \dd \kappa(b).
\end{align*}

On the other hand, the \term{external power} expended on $\Omega'$ is given by 
\[
  \Pcal(\Omega') = \int_{\partial \Omega'} t(n) \cdot \dot{y} \dd a
  + \int_{\Omega'} f \cdot \dot{y} \dd x
  - \int_{\Omega'} \rho \ddot{y} \cdot \dot{y} \dd x,
\]
where $t(n) \in \R^3$ is the surface traction in the direction of the exterior normal $n$ and $\rho > 0$ is the mass density.

\subsection{Force balance}

We now postulate, as in the general theory of continuum mechanics, the fundamental \emph{Principle of Virtual Power}~\cite[Chapter~92]{GurtinFriedAnand10book}:
\begin{quotation}
  \textit{The internal and external powers are equal, $\Ical(\Omega') = \Pcal(\Omega')$, for all allowed motions and all $\Omega' \subset \Omega$.}
\end{quotation}
Since $\dot{y}$ and $g^b$ are independent degrees of freedom, this yields
\[
  \int_{\partial \Omega'} \bigl( T n - t(n) \bigr) \cdot \dot{y} \dd a
    + \int_{\Omega'} \bigl( \rho \ddot{y} - \Diverg T - f \bigr) \cdot \dot{y} \dd x
  = 0
\]
and
\[
  - \int \int_{\Omega'}  \big(QM^Tb\big)\cdot g^b \dd x \dd \kappa(b) + \int \int_{\Omega'} X^b\cdot g^b \dd x \dd \kappa(b) = 0.
\]
Then, since $\Omega' \subset \Omega$ was arbitrary, we obtain (we drop the boundary equation since we do not discuss boundary conditions in the following)
\[
  (\rho \ddot{y} - \Diverg T) \cdot \dot{y} = f \cdot \dot{y}  \quad\text{in $\Omega$,}
\]
and
\[
  \int X^b\cdot g^b \dd \kappa(b) = \int  \big(QM^Tb\big)\cdot g^b \dd \kappa(b)  \quad\text{in $\Omega$.}
\]
As the virtual rates $\dot{y}$ and $g^b$ can attain any value, we conclude the \term{elastic force balance}
\begin{equation} \label{eq:elast_balance}
  \rho \ddot{y} - \Diverg [\DD W_e(\nabla y Q) Q^T] = f  \quad\text{in $\Omega$,}
\end{equation}
as well as the (referential) \term{plastic force balance}
\begin{equation} \label{eq:plast_balance}
  X^b =  Q M^T b =  Q \big(\DD W_e(\nabla y Q)\big)^T\nabla y Q b
  = Q \big(\DD W_e(E)\big)^T Eb\quad\text{in $\Omega$}
\end{equation}
which must hold for each $b \in \Bcal$.

We may alternatively express the plastic force balance in the structural configuration, writing
\[
  PX^b = Q^{-1} X^b = M^Tb = \big(\DD W_e(\nabla y Q)\big)^T\nabla y Q b = \big(\DD W_e(E)\big)^T Eb,
\]
which pairs the structural configurational stress $PX^b$ with the structural rate $Q^Tg^b$. Indeed, the constitutive assumptions we discuss in the following section will indeed relate the structural stress $PX^b$ to the structural rate $Q^Tg^b$. We remark that while the stress depends upon the frame in which we express it, the power does not.

We further impose the \emph{Free Energy Imbalance}~\cite[Section~27.3]{GurtinFriedAnand10book}, which is itself a consequence of the \emph{Second Law of Thermodynamics}:
\begin{quotation}
  \textit{Along the evolution it holds that $\displaystyle \frac{\dd}{\dd t}\Wcal_e(\Omega')-\Pcal(\Omega')=-\Delta(\Omega') \leq 0$ in every subregion $\Omega' \subset \Omega$.}
\end{quotation}
By the \emph{Principle of Virtual Power} we also have $\Pcal(\Omega')=\Ical(\Omega')$. Thus we obtain
\[
  \frac{\dd}{\dd t}\Wcal_e(\Omega')-\Ical(\Omega')=-\Delta(\Omega') \leq 0.
\]
Using analogous derivations like the ones for the internal power above, we obtain the \term{dissipation relation}
\begin{equation} \label{eq:thermodyn_dissip}
  \int \int_{\Omega'} X^b\cdot g^b \dd x \dd \kappa(b) = \Delta(\Omega') \geq 0,
\end{equation}
where $\Delta(\Omega') \in [0,+\infty)$ is the \term{dissipation rate} in $\Omega'$. This expresses the \emph{irreversibility} of plastic flow whenever $\Delta(\Omega') > 0$. In contrast, purely elastic deformations (for which $g^b \equiv 0$ and hence $\Delta(\Omega') = 0$) are reversible.

\subsection{Flow rule} \label{sc:flowrule}
So far we have not constitutively specified the configurational stresses $X^b$ in~\eqref{eq:plast_balance}. We propose that~\eqref{eq:plast_balance} can be written as a differential inclusion for a function of the \emph{structural} rate $Q^Tg^b$, where $g^b$ is the geometric rate. This expresses the idea that rates must be mapped into the lattice in order to evaluate the dissipated energy. Indeed, since in our model the primary source of anisotropy lies in the crystal lattice (different behavior along the crystal vectors), the structural dissipation potentials have a good chance of being \enquote{simpler} than their referential counterparts. This approach is in agreement with the formulation for phenomenological plastic evolution in much of the literature (see, e.g., ~\cite{Mielke03a,GurtinFriedAnand10book}). The precise form of this constitutive assumption is the following (see also~\cite[Section~4.4]{Fremond02} and~\cite{Silhavy97} for similar models):
\begin{quotation}
  \textit{There are \term{dissipation (pseudo)potentials} $R ^b \colon \R^3 \to [0,+\infty]$ ($b \in \Bcal$), which are proper ($\not\equiv +\infty$), convex, lower semicontinuous, and satisfy $R^b(0) = 0$ as well as the symmetry relation $R^{-b}(-\xi) = R^b(\xi)$ for all $b \in \Bcal$, such that the \term{flow rule}
\begin{equation} \label{eq:flow}
  P X^b \in \partial R^b (Q^T g^b)
\end{equation}
holds.}
\end{quotation}
Here, \enquote{$\partial R^b$} denotes the convex subdifferential of $R^b$, that is, $\partial R^b(Q^T g^b)$ consists of all those \term{(normal) flow stresses} $\sigma \in \R^3$ that satisfy
\[
  R^b (Q^T g^b) + \sigma \cdot (\xi - Q^T g^b) \leq R^b(\xi) \qquad
  \text{for all $\xi \in \R^3$.}
\]
Recall that $P X^b$ is the structural configurational stress, which is power-conjugate to the structural geometric slip rate $Q^T g^b$, so~\eqref{eq:flow} indeed pairs power-conjugate quantities. That the flow rule is formulated as a \emph{differential inclusion}, often called the \term{Biot inclusion}, is chiefly a constitutive assumption, cf.~Section~4.4 in~\cite{Fremond02} as well as~\cite{ZieglerWehrli87,Silhavy97,Mielke03a}. In principle, we can also allow for more general dissipation (pseudo)potentials, with $R^b$ depending on further quantities, such as the type of dislocation (edge or screw). For notational reasons we have not written such dependencies explicitly.

If $R^b$ is positively $k$-homogeneous ($k \geq 0$), then the convexity of $R^b$ expresses the following intuitive constraint: When the system is moving in direction $Q^T g^b =: \xi = (1-\theta) \xi_0 + \theta \xi_1 \neq 0$, where $\xi_0, \xi_1 \in \R^3 \setminus \{0\}$, $\theta \in (0,1)$, then the frictional power is $\xi \cdot \partial R^b(\xi) = k R^b(\xi)$ (element-wise, by Euler's positive homogeneity theorem). Alternatively, the rate could oscillate very quickly between rates $\xi_0, \xi_1$ with time-fractions $1-\theta$ and $\theta$, respectively, which would result in the frictional power $(1-\theta) k R^b(\xi_0) + \theta k R^b(\xi_1)$. The convexity tells us that this oscillatory path expends at least as much energy as the non-oscillatory one. In fact, if this is not satisfied (e.g.,  because of the presence of microstructure), then the material would choose the oscillatory paths in an optimal way at a small length scale. Hence, at the meso- or macroscopic length scale we would observe the \emph{effective} dissipation potential, which is a suitable relaxation of the microscopic dissipation potential; see~\cite{ContiGarroniOrtiz15} for a similar effect. Moreover, it can be shown that the \emph{Maximum Plastic Work Principle}, which is a strengthening of the \emph{Second Law of Thermodynamics}, implies convexity of $R^b$, see~\cite[pp.57--59]{HanReddy13}. The lower semicontinuity is a minimal continuity assumption and can again be justified on physical grounds (similarly to the justification of convexity).

According to the Coleman--Noll procedure~\cite{ColemanNoll63,Gurtin2000,GurtinFriedAnand10book}, thermodynamic reasoning should give \emph{constitutive restrictions}. In this spirit, combining~\eqref{eq:thermodyn_dissip} with the flow rule~\eqref{eq:flow} yields the \term{strict positivity} condition on the dissipation potential $R^b$:
\begin{quotation}
  \textit{$\xi \cdot \partial R^b(\xi) > 0$ (element-wise) for non-zero $\xi \in \R^3$.}
\end{quotation}
Thus, energy is dissipated if and only if the material flows plastically. Whenever $R^b$ is positively homogeneous of any order, then this is equivalent to $R^b(\xi) > 0$ for $\xi \neq 0$ (again by Euler's positive homogeneity theorem).

If we prescribe that $P \in \Pfrak$ for a Lie group $\Pfrak \subset \SL(3)$ as in Section~\ref{sc:restrict}, then we need to require that $QD \in \pfrak$, so that $\dot{Q}Q^{-1} \in \pfrak$ and the plastic flow does not leave $\Pfrak$. According to the definition of the total plastic drift $D$ in~\eqref{eq:D}, this is true if $Qb \otimes g^b \notin \pfrak$ implies $R^b(Q^T g^b) =+\infty$. Then, $\partial R^b(Q^T g^b) = \emptyset$, and so the flow cannot progress. Setting $\xi := Q^T g^b$, observe that $Qb \otimes g^b \notin \pfrak$ if and only if $b \otimes \xi = b \otimes (Q^T g^b) = Q^{-1}[Qb \otimes g^b]Q \notin \pfrak$ (since matrix Lie algebras are invariant under matrix similarity transformations). Thus, we arrive at the following condition, which acts as a restriction on the constitutive choice of $R^b$:
\begin{quotation}
  \textit{If $b \otimes \xi \notin \pfrak$ for $\xi \in \R^3$, then $R^b(\xi) = +\infty$.}
\end{quotation}

\begin{example} \label{ex:SLcont}
If $\Pfrak = \SL(3)$ then, as shown in Example~\ref{ex:SL}, we need that $\tr(QD) = 0$. To ensure this, the above condition reads as follows: If $R^b(Q^T g^b) < +\infty$, then $b \perp Q^T g^b$. In this case,
\[
  0 = \tr \bigl( b \otimes Q^T g^b \bigr)
  = \tr \bigl( (b \otimes g^b)Q \bigr)
  = \tr \bigl( (Qb) \otimes g^b \bigr),
\]
that is, $Qb \perp g^b$, which is the characteristic relation for dislocation glide~\cite{HullBacon11book}. Hence, the above condition on $R^b$ means that climb is forbidden.
\end{example}

Finally, we observe that the flow rule~\eqref{eq:flow} can also be written referentially as $X^b \in \partial \hat{R}^b(Q,g^b)$, where the subdifferential is taken in the second variable, and $\hat{R}^b(Q,g^b) := R^b(Q^Tg^b)$. Furthermore, if we wish to formulate the problem in terms of the $2$-vector-valued geometric rate $\gamma^b$ (defined in Section~\ref{sc:plastflow_rigorous} as the Hodge dual of $g^b$) and a conjugate stress $\tilde{X}^b\in \Wedge^2\R^3$, we can do so by defining an alternate dissipation potential $\tilde{R}^b:\Wedge_2\R^3\to[0,+\infty]$ via $\tilde{R}^b(\xi) = R^b(\hodge \xi)$ (where $\star$ is the Hodge star defined in Section~\ref{sc:algebra}) for any $\xi\in \Wedge_2\R^3$. The flow rule then becomes
\begin{equation} \label{eq:flowrule2vec}
  Q^T \tilde{X}^b\in \partial \tilde{R}^b(P\gamma^b).
\end{equation}
To explain the relationship between $P\gamma^b$ and $Q^Tg^b$, we recall that $g^b$ is the oriented unit normal to the plane defined by the $2$-vector $\gamma^b$, as defined in~\eqref{eq:gammab}. When considering the dissipation caused by dislocation motion represented by $\gamma^b$, it is natural to transform this plane into the structural configuration using the inverse of the scaffold map, $P = Q^{-1}$, so that we consider any motion within the undistorted lattice. However, as $Q$ need not be an orthogonal transformation, the normal vector to the resulting structural plane is not simply transformed by $P$. Instead, we find that if $\gamma^b\mapsto P\gamma^b$, then $g^b\mapsto P^{-T}g^b = Q^Tg^b$ (and $Q^T \neq P$ if $P$ is not orthogonal). A full argument explaining this formula is given in Section~\ref{sc:slipderivation}.

\subsection{Duality and stability}

One may further define the \term{dual dissipation potential} $R^{b*} \colon \R^3 \to [0,+\infty]$ as the convex conjugate function of $R^b$, that is,
\[
  R^{b*}(\sigma) := \sup\, \setb{ \dpr{\xi,\sigma}  - R^b(\xi)}{ \xi \in \R^3 },  \qquad \sigma \in \R^3.
\]
Then, by standard results in convex analysis, see for example~\cite{Rock70CA}, we have that $R^{b*}$ is also proper, convex, lower semicontinuous, and satisfies the same symmetry relation as the $R^b$. Moreover, the flow rule~\eqref{eq:flow} is equivalent to the \term{dual flow rule} (\term{Onsager inclusion}),
\begin{equation} \label{eq:flow_dual}
  Q^Tg^b \in \partial R^{b*}(P X^b).
\end{equation}
This follows from the Legendre--Fenchel theorem and standard rules for the computation of convex conjugates.

It is a fundamental property of elasto-plastic processes that if the (structural) configurational stress $PX^b$ lies in the interior of an \term{elastic stability domain} $\Scal^b \subset \R^3$ ($b \in \Bcal$), then no plastic flow takes place. The thinking here is that below the stress threshold $\partial \Scal^b$, which is called the \term{yield surface}, no plastic slip can be activated. This is in very good agreement with experiments, see Chapter~5 in~\cite{LemaitreChaboche90}. We set
\[
  \Scal^b := \partial R^b(0) \subset \R^3.
\]
It follows from standard results in convex analysis that $\Scal^b$ is a closed, convex neighborhood of the origin and $\Scal^{-b} = -\Scal^b$.

Define the \term{rate-independent dissipation potential} $R^b_1 \colon \R^3 \to [0,+\infty]$ as the support function of $\Scal^b$, i.e., the dual of the characteristic function $\chi_{\Scal^b}$ of $\Scal^b$ (which is $0$ on $\Scal^b$ and $+\infty$ otherwise),
\[
  R^b_1(\xi) := \chi_{\Scal^b}^*(\xi) = \sup_{\sigma \in \Scal^b} \dprb{\xi, \sigma}.
\]
We note that $R^b_1 \geq 0$ since $0 \in \Scal^b$. Moreover, let the \term{residual dissipation potential} $R^b_+ \colon \R^3 \to [0,+\infty]$ be defined as $R^b_+ := R^b - R^b_1$, so that we have the splitting
\[
  R^b(\xi) = R^b_1(\xi) + R^b_+(\xi),  \qquad \xi \in \R^3.
\]
It can be seen by arguing via supporting hyperplanes of the epigraph of $R^b_+$ that $R^b_+$ is proper, convex, lower semicontinuous, non-negative, $R^b_+(0) = 0$, and the same symmetry relation as for $R^b$ holds.

Using the inf-convolution $(f \infc g)(x) := \inf_z [f(x-z)+g(z)]$ for proper, convex, lower semicontinuous $f,g$, and the associated duality rule $[f \infc g]^* = f^* + g^*$ (see Theorem~16.4 of~\cite{Rock70CA}), we compute
\[
  R^{b*}
  = [R^b_1 + R^b_+]^*
  = [R^b_1]^* \infc R^{b*}_+
  = \inf_{\sigma \in \Scal^b} R^{b*}_+(\frarg-\sigma).
\]
In particular,
\[
  R^{b*}(P  \sigma^b) = 0  \quad\text{if}\quad
  P \sigma^b \in \Scal^b.
\]
Consequently,
\[
  \partial R^{b*}(P  \sigma^b) = \{0\}  \quad\text{if}\quad
  P  \sigma^b \in (\Scal^b)^\circ,
\]
where $(\Scal^b)^\circ := \Scal^b \setminus \partial \Scal^b$ is the interior of $\Scal^b$. This expresses that the flow stops once the stress attains a value in the interior the elastic stability domain $\Scal^b$, as required.

\subsection{Dissipation} \label{sc:diss}

For the dissipation one can consider two main possibilities:
\begin{enumerate}[(i)]
  \item \emph{The rate-dependent case:} $R^b_+(\xi) > 0$ for all $\xi \neq 0$.
  \item \emph{The rate-independent case:} $R^b_+ \equiv 0$.
\end{enumerate}
The former case corresponds to modeling situations for which the rate of loading and of dislocation motion are comparable, while the latter corresponds to the case where dislocation motion is much more rapid than the loading rate. In this context we also refer to~\cite{Hudson17} for an example where a rate-dependent dissipation potential for dislocation motion is derived from first principles.

\begin{example} \label{ex:powerlaw}
In rate-dependent plasticity theory it is common to choose a power law for the dissipation potential $R^b \colon \R^3 \to [0,\infty)$, see for instance~\cite[Section~101]{GurtinFriedAnand10book}. For this, set for $\kappa > 0$ and $N > 1$,
\[
  R^b(\xi) := \frac{\kappa}{1+1/N} \abs{\xi}^{1+1/N},  \qquad \xi \in \R^3 ,
\]
where $\abs{\frarg}$ is a vector norm (e.g., the Euclidean norm in the isotropic case). Then, $\DD R^b(\xi) = \kappa\abs{\xi}^{1/N}$ and also
\[
  R^{b*}(\sigma) = \frac{\kappa^{-N}}{1+N} \abs{\sigma}_*^{1+N},  \qquad \sigma \in \R^3,
\]
with $\abs{\frarg}_*$ the dual norm to $\abs{\frarg}$.
\end{example}

We define the \term{(total) dissipation} over an interval $[s,t]$ as
\begin{equation} \label{eq:Diss}
  \Diss([s,t])
  := \int_s^t \Delta(\tau) \dd \tau
  = \int_s^t \int \int_{\Omega} X^b\cdot g^b \dd x \dd \kappa(b) \dd \tau
  \geq 0.
\end{equation}
Now use the Legendre--Fenchel theorem to obtain the splitting
\begin{align*}
  \Diss([s,t])
  &= \int_s^t \int \int_{\Omega} X^b\cdot g^b \dd x \dd \kappa(b) \dd \tau \\
  &= \int_s^t \int \int_{\Omega} P X^b\cdot Q^Tg^b \dd x \dd \kappa(b) \dd \tau \\
  &= \int_s^t \int \int_{\Omega} R^b(Q^T g^b) + R^{b*}(P X^b) \dd x \dd \kappa(b) \dd \tau \\
  &= \int_s^t \int \int_{\Omega} R^b_1(Q^T g^b) + R^b_+(Q^T g^b) + R^{b*}(P X^b) \dd x\dd \kappa(b) \dd \tau \\
  &= \Diss_1([s,t]) + \Diss_+([s,t])
\end{align*}
with
\begin{align*}
  \Diss_1([s,t])
  &:= \int_s^t \int \int_{\Omega} R^b_1(Q^T g^b) \dd x \dd \kappa(b) \dd \tau
\end{align*}
and
\begin{align*}
  \Diss_+([s,t])
  &:= \int_s^t \int \int_{\Omega} R^b_+(Q^T g^b) + R^{b*}(P X^b) \dd x \dd \kappa(b) \dd \tau.
\end{align*}

In the rate-dependent case, we have $P X^b \in \partial R^b_1(Q^T g^b) + \DD R^b_+(Q^T g^b)$. Furthermore, we use Euler's homogeneity theorem to see that $(Q^Tg^b)\cdot \partial R^b_1(Q^T g^b) = R^b_1(Q^T g^b)$ (element-wise). Then, $P X^b\cdot Q^T g^b = R^b_1(Q^T g^b) + (Q^T g^b)\cdot\DD R^b_+(Q^T g^b)$ and so
\[
  \Diss_+([s,t]) = \int_s^t \int \int_{\Omega} (Q^T g^b)\cdot\DD R^b_+(Q^T g^b) \dd \kappa(b) \dd \tau.
\]
In the rate-independent case, $R^{b*} = \chi_{\Scal^b} = R^{b*}_1$ so $\Diss_+([s,t]) = 0$.

\subsection{Linearization and Peach--Koehler force} \label{sc:PKstress}

In this final section we demonstrate that linearizing our formulation yields the classical expressions of linear elasto-plasticity.

We suppose that the total deformation gradient is expressible as a perturbation of the identity, i.e., $y(t,x) = x+ u(t,x)$, where $u = u(t,x) \in \R^3$ is the displacement. We assume further that the displacement gradient $\nabla u$ is uniformly small, so that the deformation gradient is a perturbation of the identity matrix, $\nabla y=\Id + \nabla u$. Likewise, we assume that the plastic distortion can also be expressed as a perturbation of the identity, $P =\Id + Z$, where $Z$ is again uniformly small (and of a comparable magnitude to the displacement gradient $\nabla u$). The linearized form of the plastic flow equation~\eqref{eq:plast_flow} becomes
\[
  \dot{P} = \dot{Z} = D.
\]
Considering the elastic strain next, we note that $E$ can be formally expanded and approximated as
\[
  E = \nabla y Q = (\Id + \nabla u)(\Id + Z)^{-1} = \Id + \nabla u-Z -\nabla u\, Z +Z^2+\cdots\approx \Id + \nabla u -Z.
\]
This leads to the usual definition of the linearized elastic distortion $\beta_e := \nabla u -Z$. Taylor-expanding the Piola--Kirchoff stress, and assuming that the state of zero elastic strain is stress free, i.e., $\DD W_e(\Id) = 0$, we define the fourth-order elastic tensor $\mathsf{C}:=\DD^2W_e(\Id)$. Then, retaining only the leading-order terms, we obtain that
\[
  T = \DD W_e(E)Q^T = \DD W_e(\Id + \beta_e)(\Id + Z)^{-T} \approx (\DD^2 W_e(\Id):\beta_e)(\Id + Z)^{-T} \approx \mathsf{C}:\beta_e.
\]
As a result, the linearized form of the elastic force balance~\eqref{eq:elast_balance} becomes
\[
  \rho \ddot{u} - \Diverg [\mathsf{C}:\beta_e] = f.
\]
Applying similar considerations to the expression for the Mandel stress in~\eqref{eq:Mandel}, we have
\[
  M = E^T \DD W_e(E) = (\Id + \beta_e)^T \DD W_e(\Id + \beta_e) \approx (\Id + \beta_e)^T \mathsf{C}:\beta_e \approx \mathsf{C}:\beta_e.
\]
and so the linearized Mandel stress is identical to the linearized Piola--Kirchhoff stress, in contrast to the fully nonlinear setting.

Early in the development of the theory of dislocations, the force acting on a dislocation line within a linear theory of elasto-plasticity was derived through a themodynamical argument~\cite{PeachKoehler50}. Neglecting line tension and terms that are quadratic in $\beta_e$ or $Z$, the configurational stress $X^b$ can be approximated as follows:
\[
  X^b = QM^Tb = (\Id+Z)^{-1}M^Tb \approx (\Id-Z+\dots)M^Tb \approx (\mathsf{C}:\beta_e)^T b.
\]
We can express the pairing between the configurational stress $X^b$ and the geometric rate $g^b$ to approximate the rate of power expended by a moving dislocation as
\[
  X^b \cdot g^b \approx \bigl( (\mathsf{C}:\beta_e)^T b \bigr) \cdot \bigl(\eta *\big[ v^b\times \vec{T}^b\, m^b\big]\bigr).
\]
The Peach--Koehler force~\cite{PeachKoehler50} was formally derived as the force which acts to oppose the motion of an infinitesimally thin dislocation, so that the power expended is equated with the negative of the work done by this force,
\[
  X^b \cdot g^b = -f^b\cdot v^b.
\]
Formally setting $\eta$ to be a Dirac delta, and $m^b=1$ to reflect the case of a single dislocation, and manipulating the expression above using the permutation invariance of the scalar triple product, we have
\[
  \bigl((\mathsf{C}:\beta_e)^T b\bigr) \cdot \bigl[ v^b\times \vec{T}^b \bigr]=
  v^b\cdot\bigl[\vec{T}^b \times \bigl((\mathsf{C}:\beta_e)^T b\bigr) \bigr] = -v^b\cdot\bigl[ \bigl((\mathsf{C}:\beta_e)^T b\bigr) \times \vec{T}^b  \bigr],
\]
and so we find through this linearization process that we obtain the classical definition of the Peach-Koehler force, namely
\[
    f^b \approx \bigl((\mathsf{C}:\beta_e)^T b\bigr) \times \vec{T}^b.
\]

\section{Summary of model} \label{sc:summary}

In the following we summarize the relations making up our model.

\begin{enumerate}[(1)] \setlength\itemsep{8pt}
\item Fundamental variables: $y = y(t,x) \in \R^3$ the deformation and $Q = Q(t,x) \in \SL(3)$ the crystal scaffold, or, equivalently, $P=Q^{-1}$ the plastic distortion.

\item Kr\"{o}ner decomposition: $\nabla y = EQ^{-1} = EP$, where $E := \nabla y Q$ is the elastic distortion.

\item Plastic restriction: $Q \in \Pfrak$ for a Lie group $\Pfrak \subset \SL(3)$ with Lie algebra $\pfrak = \Lie(\Pfrak)$.

\item Burgers measure: $\kappa := \frac12 \sum_{b \in \Bcal} \delta_b$.
\item Slip trajectories: A collection of two-dimensional surfaces in space-time, $\Sigma = (S^b)_{b\in\Bcal}$, with
\[
  S^{-b} = -S^b, \qquad
  \partial S^b \restrict ((0,T) \times \Omega) = 0.
\]

\item Dislocation system at time $t$: $(T^b(t))_b$ with $T^b(t) := \pbf_*(S^b|_t)$.

\item Thickened slip trajectories $\Sigma_\eta := (S^b_\eta)_b$, where $S^b_\eta := \eta \conv S^b$ with $\eta$ a smooth and compactly supported dislocation line profile $\eta \colon \R^3 \to [0,\infty)$ satisfying $\int \eta \dd x = 1$.

\item Geometric slip rate:
\[
  g^b(t,\frarg)
  := \eta \conv \biggl[ \frac{\DD}{\DD t} S^b(t,\frarg) \times \vec{T}^b(t,\frarg) \, m^b \biggr].
\]
with $m^b$ the multiplicity of $T^b(t)$ and the dislocation velocity
\[
  \frac{\DD}{\DD t} S^b(t,x) := \frac{\pbf(\xi^b(t,x))}{\abs{\tbf(\xi^b(t,x))}},
\]
where
\[
  \xi^b(t,x) := \frac{\nabla^{S^b} \tbf(t,x)}{\abs{\nabla^{S^b} \tbf(t,x)}}.
\]
This $\xi^b$ can equivalently be defined as the unique vector such that $\Trm_{(t,x)}S^b = \spn\{\vec{T}^b(t,x),\xi^b(t,x)\}$ and such that $\xi^b(t,x)$ is pointing \enquote{forward in time}, i.e., $\tbf(\xi^b(t,x)) = \xi^b(t,x)\cdot \ee_0\geq 0$.

\item Regularity (no horizontal pieces): $\tbf(\xi^b(t,x)) = \xi^b(t,x)\cdot \ee_0 > 0$.
  
\item Plastic flow equation:
\[
  \dot{P} = -Q^{-1} \dot{Q} Q^{-1} = D,
\]
where the total plastic drift is given as
\[
  D(t,x) := \int b \otimes g^b(t,x) \dd \kappa(b).
\]

\item Total energy:
\[
  \Ecal(t,y,Q,\Phi) := \Wcal_e(y,Q) - \int f(t) \cdot y \dd x,
\]
where $f \colon [0,T] \times \Omega \to \R^3$ is the external bulk loading and the elastic energy is given as
\begin{equation*}
  \Wcal_e(y,Q) := \int_\Omega W_e(\nabla y Q) \dd x,
\end{equation*}
with $W_e \colon \GL^+(3) \to \R$ the (frame-indifferent) elastic energy density.

\item Piola--Kirchhoff stress (referential elastic stress), Mandel stress (structural plastic stress), and referential configurational stress:
\begin{align*}
  T &:= \DD W_e(\nabla y Q) Q^T,  \\
  M &:= Q^T \nabla y^T \DD W_e(\nabla y Q),  \\
  X^b &:= Q M^T b,  \qquad b \in \Bcal.
\end{align*}

\item Elastic force balance:
\[
  \rho \ddot{y} - \Diverg [\DD W_e(\nabla y Q) Q^T] = f.
\]

\item Flow rule (Biot inclusion):
\[
  P X^b \in \partial R^b(Q^T g^b),  \qquad b \in \Bcal,
\]
with the dissipation potential $R^b \colon \R^3 \to [0,+\infty]$ satisfying:
\begin{align*}
  &\text{$R^b$ is proper ($\not\equiv +\infty$), convex, lower semicontinuous, $R^b(0) = 0$;} \\
  &\text{the symmetry relation $R^{-b}(-\xi) = R^b(\xi)$ holds for all $b \in \Bcal$;} \\
  &\text{$\xi \cdot \partial R^b(\xi) > 0$ (element-wise) for non-zero $\xi \in \R^3$;} \\
  &\text{If $b \otimes \xi \notin \pfrak$ for $\xi \in \R^3$, then $R^b(\xi) = +\infty$.}
\end{align*}

\end{enumerate}

\section{Further effects and outlook}\label{sc:outlook}

In this section we outline how further physical effects may be incorporated into our model, and describe various directions for future work.

\subsection{Elastic equilibrium} \label{sc:elast_equib}

Let us consider the consequences of imposing the following further assumption:
\begin{quotation}
  \textit{Inertial effects can be neglected.}
\end{quotation}
This is a common hypothesis in both macroscopic plasticity modeling and many micromechanical approaches to dislocation motion. The main theoretical justification for assuming this is that the frictional forces for dislocation motion dominate inertial effects in any slow-loading process~\cite{BulatovCai06book}.

In order to formulate the resulting equations, we further need a qualitative relationship between elastic and plastic motion, for which we assume:
\begin{quotation}
  \textit{Elastic relaxation occurs on a much faster timescale than plastic flow.}
\end{quotation}
Experiments suggest that this is a reasonable assumption in many circumstances, see for example~\cite{ArmstrongArnoldZerilli09,BenDavidEtAl14}. It is also theoretically consistent (at least for metals) since the flow of dislocations is constricted to be always slower than the propagation of elastic distortion through shear waves (S-waves), because plastic drag tends to infinity as the plastic distortion rate approaches the shear wave speed; see~\cite{DeHossonyRoosMetselaar02}. In this case, we obtain that the \term{elastic equilibrium equation} (where the inertial term has been dropped)
\[
  - \Diverg T = - \Diverg [ \DD W_e(\nabla y Q) Q^T ] = f
\]
is always satisfied, even during plastic flow. In other words, the evolution is \emph{quasi-static} (see, e.g.,~\cite{MielkeRoubicek15book} for quasi-static evolution). If the material is hyperelastic, then this is equivalent to $y(t)$ being a minimizer of $\Ecal(t,\frarg,Q(t))$ for all $t$.

\subsection{Dislocation climb} \label{sc:climb}
If we wanted to dispense with the plastic incompressibility assumption in order to allow climb as well as glide, then we need to define the total plastic drift in~\eqref{eq:D} instead as follows:
\[
  D(t,x) := \int b \otimes \proj_{\langle Q(t,x)b \rangle^\perp} g^b(t,x) \dd \kappa(b),
\]  
where by $\proj_{\langle Q(t,x)b \rangle^\perp}$ we denote the orthogonal projection onto the orthogonal complement of the line $\langle Q(t,x)b \rangle = \spn \{Q(t,x)b\}$. This ensures that $\tr(QD) = 0$ (see Examples~\ref{ex:SL} and~\ref{ex:SLcont}) and so the scaffold $Q$ remains in $\SL(3)$; without this restriction, it is unclear how to interpret the scaffold $Q$ once $\det Q \neq 1$. A natural way to account for the volumetric change caused by climb would be to add a further field describing point defects.

\subsection{Rate-independent evolution} 

Experiments show that plastic flow is \emph{rate-dependent} (viscous), but only slightly so below absolute temperatures of approximately $0.35 \vartheta_m$, where $\vartheta_m$ is the melting temperature of the material, see~\cite[Section~78]{GurtinFriedAnand10book}. For instance, in a commonly-used power viscosity law, see Example~\ref{ex:powerlaw} and also~\cite[Section~5.4]{LemaitreChaboche90}, the stress depends on the rate with exponent $1/N$, that is, $\DD R(\xi) \sim \abs{\xi}^{1/N}$. For example, steel with 35\% carbon at 450 \textdegree{}C has $N = 15$ and the titanium-aluminium alloy TA6V at 350 \textdegree{}C has $N = 120$; more values can be found in~\cite[Table~6.2]{LemaitreChaboche90}. The movement of the system is directed in such a way as to move the stress toward the yield surface, where the movement stops as the internal friction stress threshold is no longer exceeded. It can be seen that the larger $N$ is, the faster this \enquote{plastic relaxation} takes place. 

On \enquote{slow} time scales we should therefore see near-infinitely fast relaxation, i.e.\ $PX^b \in \Scal^b$ always, which is called \emph{(local) stability}. If $f$ were to be held constant at some point in time, the system would settle very quickly into a rest state until the external loading changes and the system is pushed out of equilibrium. The traditional rate-independent modeling is built upon the assumption that only this global movement is interesting and the fast \enquote{relaxation} movements towards a rest state can be neglected, at least if the system does not jump to a far-away state in an instant.

A further key question, which is also central in other recent works~\cite{OrtizRepetto99,Mielke02,Mielke03a,MielkeRossiSavare09,MielkeRossiSavare12,DalMasoDeSimoneSolombrino10,DalMasoDeSimoneSolombrino11,RindlerSchwarzacherVelazquez21}, is whether during a jump at \enquote{infinite} speed the modeling assumption of rate-independence can be upheld. Most materials in fact display rate-dependent behavior under fast deformations.

The work~\cite{Rindler21b?} will implement a rate-independent version of the model presented here, which relies on a number of further technicalities and a reformulation of the flow rule into a stability inequality and an energy balance.

\subsection{Core energy} \label{sc:core_energy}

To incorporate the fact that the energy of a dislocation may not exclusively be captured via the elastic energy, one may also add a \term{core energy} for the dislocation system $\Phi = (T^b)_b$ to the total energy. This expresses an additional chemical potential energy in the system caused by the fact that dislocation configurations are energetically unfavourable in comparison to the undistorted lattice on an \emph{atomistic} level. It does \emph{not} model the elastic potential energy \enquote{trapped} in the elastic distortion, which cannot be released since $\curl P \neq 0$ at a dislocation (this effect is included automatically).

A simple model for the core energy is
\[
  \Wcal_c(\Phi) := \zeta \int \Mbf(T^b) \dd \kappa(b),
\]
where $\zeta > 0$ is a material constant and $\Mbf(T^b)$ is the total length of the dislocations with Burgers vector $b$. More complicated expressions (e.g.,  with anisotropy or $b$-dependent $\zeta$) are of course possible, as in~\cite{ArizaContiGarroniOrtiz18}.

Using the simple core energy above, the total energy is then modified from~\eqref{eq:E} to
\[
  \Ecal(t,y,Q,\Phi) := \Wcal_e(y,Q) - \int f(t) \cdot y \dd x + \Wcal_c(\Phi)
\]
and the plastic force balance~\eqref{eq:plast_balance} would also need to incorporate a \emph{curvature}-type term.

\subsection{Hardening and softening}

In its most general form, hardening or softening describe the processes by which the \emph{effective} elastic stability domain changes, in particular expands or contracts, due to a change in the internal state of the specimen. Hardening is usually anisotropic and due to a variety of microscopic effects like dislocation entanglement. We discuss some connections to existing models and approaches here.

One simple way to add hardening to our model is to add a prefactor to the dissipational cost \enquote{$\Diss$}, which depends on $P$ or further internal variables. This is the approach taken in~\cite{Rindler21b?} and we refer to that work for one possible way to implement hardening effects that yields a mathematically well-posed theory.

Another, more classical, way is to add a \enquote{hardening energy} to the total energy, which depends on further internal variables. Then, one encounters another (generalized) stress, which is power-conjugate to the rate of change in the internal variables. We denote the state space of the additional internal variables by $\Zfrak$ and the corresponding state by $z = z(t,x)$. For simplicity we assume that $\Zfrak$ has a linear structure, so that $\dot{z}$ is the rate of change for $z$ (if $\Zfrak$ had a Lie group structure, we could mimick the previous development for $P$ and consider the internal variable drift $\dot{z}z^{-1} $ in the structural frame or $z^{-1} \dot{z}$ in the referential frame; see~\cite{Mielke02} for more on this approach). We then add a \term{hardening energy}
\[
  \Wcal_h(z) := \int_\Omega W_h(z(x)) \dd x
\]
with $W_h \colon \Zfrak \to [0,\infty)$, to our total energy. The flow rule now involves the plastic as well as the internal rates and stresses, and hence the signature of $R^b$ has to be suitably adapted.

In \term{isotropic hardening}, the elastic stability domain $\Scal^b$ ($b \in \Bcal$) remains centered around the origin but can expand in what is called \term{positive hardening} and contract in \term{negative hardening} or \term{softening}. In \term{kinematic hardening}, the elastic stability domain is translated. Combined, these two effects give a first approximation to the often-observed phenomenon that an increase in tensile yield strength goes along with a decrease in compressive yield strength, called the \term{Bauschinger effect}, which is in general more complex, see for instance~\cite[Section~3.3.7]{LemaitreChaboche90} and~\cite{KassnerEtAl09}.

\begin{example} \label{ex:vonMises_istrop_kinemat}
In the often-considered \term{von Mises isotropic--kinematic hardening}, the elastic stability domain $\Scal^b$ depends on two internal variables, $H^b \in \pfrak^* \subset \R^{3 \times 3}$ and $\zeta^b \in \R$, called \term{backstresses}. Then, the elastic stability domain is prescribed to be
\[
  \Scal^b = \setB{ \sigma=(\sigma^b,H^b,\zeta^b) }{ \abs{\dev(\sigma^b-H^b)} + \zeta^b - \textstyle\frac{2}{3} \sigma^b_0 \leq 0 },
\]
where $\sigma^b_0 > 0$ is a constant (the initial tensile yield strength) and, as usual, $\abs{A} = \abs{A}_F = [\tr (A^T A)]^{1/2}$ is the Frobenius norm. This means that the elastic stability domain is translated with $H^b$ and dilated with $\zeta^b$. Using a different matrix norm, one can get different shapes of the yield surface.
\end{example}

Other hardening models (e.g.,  Tresca, Mohr--Coulomb or Drucker--Prager) could likewise be incorporated; we refer to~\cite[p.66~ff.]{HanReddy13} for descriptions of these effects.

\subsection{Coarse-graining}

Our model falls into the category of \enquote{semi-discrete} dislocation models, which occupy a position on length-scales above fully discrete lattice models like those studied in~\cite{ArizaOrtiz05,Ponsiglione07,HudsonOrtner14,ADLGP14,HudsonOrtner15,Hudson17,ADLGP17,Williams17PhD,HvMP20} since we have effectively let the lattice spacing tend to zero. It would be interesting to investigate if this relationship can be made rigorous.

On the other hand, our model works with length-scales below typical plasticity models since we still account for individual dislocation lines. One advantage of the mathematical machinery which comes with using currents is that is also immediately suggests methods to enable passage from a model representing individual dislocation lines to \emph{fields} of dislocation lines. As real materials usually contain huge numbers of dislocation lines, with total length per unit volume usually around $10^{10} \,\mathrm{m}^{-2}$ to $10^{12} \,\mathrm{m}^{-2}$ in well-annealed crystals and up to $10^{15} \,\mathrm{m}^{-2}$ after heavy plastic distortion, see~\cite{HullBacon11book}, it is a sensible mathematical abstraction to pass to a coarse-grained continuum of lines. The latter fields are representable by \emph{normal currents}~\cite{Federer69book,KrantzParks08book}. Normal currents have slices and a boundary operator, so our modeling generalizes with very few modifications. Models for fields of lines already exist~\cite{Acharya01,Acharya03,AroraAcharya20,Acharya21} and it would be interesting to study which of these are selected in a suitable limit process. Also, the relationship with various results on the coarse-graining of dislocation models in lower dimensional settings~\cite{GarroniLeoniPonsiglione10,DLGP12,MSZ14,vMMP14,SPPG14,GvMPS16,CarrilloEtAl20} seems a valuable direction for future research.

The present framework also allows for the creation of new dislocation lines from point defects, which we assume to be so frequent that this creation can essentially appear anywhere (recall that in our model, dislocations are thought to sit in a mesoscale between the atomistic and macroscopic scales). One could add a point defect density field to make the model more precise in this regard.

\section{Rigorous geometric setting} \label{sc:rigorous}

In this section we rigorously define some notions and give a more precise treatment of the geometry of dislocations and slip trajectories. We also prove (still on a mostly formal level) some statements used in our derivation. A fully rigorous treatment can be found in~\cite{Rindler21a?,Rindler21b?}.

\subsection{Linear and multilinear algebra} \label{sc:algebra}

As usual, we equip $\R^3$ with its standard inner product, $u\cdot v = \sum_i u^i v^i$ where $u^i$, $v^i$ are the components of the vectors $u,v \in \R^3$. We also employ the standard orthonormal basis for this space, so that $\R^3=\spn\{\ee_1,\ee_2,\ee_3\}$. A key aspect of our approach is to consider the motion of dislocations in (Galilean) space-time, which we represent as $\R \times \R^3\cong \R^{1+3}$. This space is again endowed with the usual inner product and its canonical orthonormal basis so that $\R^{1+3}=\spn\{\ee_0,\ee_1,\ee_2,\ee_3\}$, where we abuse notation by identifying $\ee_1$, $\ee_2$ and $\ee_3$ with their natural extensions to this space, and $\ee_0 = (1,0,0,0)$ is the additional basis vector pointing in the (positive) time direction. It will be convenient to denote the orthogonal projection onto the \enquote{time} component by $\tbf \colon \R^{1+3} \to \R \times \{0\}^3 \cong \R$ and the orthogonal projection onto the \enquote{space} component by $\pbf \colon \R^{1+3} \to \{0\} \times \R^3 \cong \R^3$. We also write the natural linear extensions of these projections to multi-vectors (which we define below) using the same symbols.

The space of matrices $\R^{m \times n}$ comes with the Frobenius inner product
\[
  A : B := \sum_{i,j} A^i_j B^i_j = \tr(A^TB) = \tr(B^TA),
\]
where upper indices indicate rows and lower indices indicate columns. As matrix norm we use the induced Frobenius norm, i.e., $\abs{A} := (A : A)^{1/2} = [\tr(A^T A)]^{1/2}$.

Throughout all of the following, let $k = 0,1,2,\ldots,n$. We recall that the \term{exterior} or \term{wedge product}, denoted $\wedge$, is an anti-commutative algebraic operation which is used to study the geometry of subspaces of a vector space. In particular, the linear combinations of the $k$-fold products of vectors from $\R^n$ form the vector space of \term{$k$-vectors} $\Wedge_k \R^n$. Its dual space, the set of \term{$k$-covectors}, is $\Wedge^k \R^n$ and the duality product between these spaces is denoted by $\dpr{\frarg,\frarg} \colon \Wedge_k \R^n\times \Wedge^k \R^n\to\R$. We recall that it is common to identify $\Wedge_0 \R^n = \Wedge^0 \R^n = \R$ with the real scalars. Given the canonical bases $\{\ee_1,\ee_2,\ee_3\}$ of $\R^3\cong\Wedge_1\R^3$ and $\{\ee_0,\ee_1,\ee_2,\ee_3\}$ of $\R^{1+3}\cong\Wedge_1\R^{1+3}$, we denote the natural dual bases of $\Wedge^1\R^3$ and $\Wedge^1\R^{1+3}$ as $\{\di x^1,\di x^2,\di x^3\}$ and $\{\di x^0,\di x^1,\di x^2,\di x^3\}$, respectively, which are characterized via the relations
\[
  \dprb{\ee_i,\di x^j}
  = \delta_{ij} =
  \begin{cases}
    1 & \text{if $i=j$,}\\
    0 & \text{otherwise.}
  \end{cases}
\]
Given the distinguished nature of the time direction in our modeling, we will also write $\di t := \di x^0$.

A $k$-vector $\eta \in \Wedge_k \R^n$ is called \term{simple} if it can be written as a single $k$-fold wedge product, $\eta = v_1 \wedge \cdots \wedge v_k$ with $v_\ell \in \R^n$, and likewise for $k$-covectors. The spaces of vectors and covectors have inner products which arise naturally from the inner product of the underlying space. For simple $k$-vectors $\eta=v_1\wedge\ldots\wedge v_k$ and $\xi=w_1\wedge\ldots\wedge w_k$ this inner product is
\[
  (\eta, \xi) = \det\left(
    \begin{array}{ccc}
      v_1\cdot w_1 & \cdots & v_1\cdot w_k\\
      \vdots &\ddots & \vdots \\
      v_k\cdot w_1 &\cdots &v_k\cdot w_k
    \end{array}\right),
\]
which is extended to general $k$-vectors by linearity. A similar construction applies for $k$-covectors.

As norms on the spaces $\Wedge_k\R^n$ and $\Wedge^k\R^n$ we use the so-called \term{mass} and \term{comass norms} of $\eta\in \Wedge_k \R^n$ and $\alpha \in \Wedge^k \R^n$ instead of the norms induced by the inner products defined above. These norms are
\begin{align*}
  \abs{\eta} &:= \sup \setb{ \absb{\dprb{\eta,\alpha}} }{ \alpha \in \Wedge^k V,\;\abs{\alpha} = 1 },  \\
  \abs{\alpha} &:= \sup \setb{\absb{\dprb{\eta,\alpha}} } { \eta \in \Wedge_k V \text{ simple, unit} }.
\end{align*}
Here, a simple $k$-vector $\eta = v_1 \wedge \cdots \wedge v_k$ is called a \term{unit} if $\{v_\ell\}_\ell$ can be chosen as an orthonormal system.

For $\eta \in \Wedge_k \R^n$ and $\alpha \in \Wedge^l \R^n$ we further define the \term{restriction operators} $\eta \intprod \alpha \in \Wedge^{l-k} \R^n$ and $\eta \restrict \alpha \in \Wedge_{k-l} \R^n$ via
\begin{align*}
  \dprb{\xi,\eta \intprod \alpha} := \dprb{\xi \wedge \eta, \alpha},  \qquad \xi \in \Wedge_{l-k} \R^n,\\
  \dprb{\eta \restrict \alpha, \beta} := \dprb{\eta, \alpha \wedge \beta},  \qquad \beta \in \Wedge^{k-l} \R^n.
\end{align*}

A simple $k$-vector $\eta \in \Wedge_k \R^n$ represents an oriented $k$-plane in $\R^n$. For instance, $\eta = v_1 \wedge \cdots \wedge v_k \in \Wedge_k \R^n$ can be thought of as the $k$-plane $R := \spn \eta := \spn\{v_1,\ldots,v_k\}$ together with the orientation induced by the ordering of $v_1,\ldots,v_k$ (any other basis of $R$ has the same orientation if and only if the change-of-base matrix has positive determinant). Another way to describe an oriented $k$-plane is by providing a unit \enquote{normal} $(n-k)$-plane $S := \spn \xi$ with $\xi \in \Wedge_{n-k} \R^n$ and setting $R := S^\perp$ (with an \enquote{orthogonal orientation}). These two approaches are related via Hodge duality as follows: For a $k$-vector $\eta \in \Wedge_k \R^n$ in the cases $n=3$ or $n=4$ we consider here, we define the \term{Hodge dual} $\hodge \eta \in \Wedge_{n-k} \R^n$ as the unique vector satisfying, respectively,
\[
  \xi \wedge \hodge \eta = (\xi, \eta) \, \ee_1 \wedge \ee_2 \wedge \ee_3\qquad\text{or}\qquad
  \xi \wedge \hodge \eta = (\xi, \eta) \, \ee_0 \wedge \ee_1\wedge\ee_2 \wedge \ee_3
\]
for all $\xi \in \Wedge_k \R^n$. For $k$-covectors $\alpha \in \Wedge^k \R^n$ we may use the same formulae to define $\hodge \alpha \in \Wedge^{n-k} \R^n$, with $\ee_1 \wedge \ee_2 \wedge \ee_3$ replaced by $\di x^1\wedge\di x^2\wedge\di x^3$ and $\ee_0 \wedge \ee_1\wedge \ee_2 \wedge \ee_3$ replaced by $\di t \wedge \di x^1\wedge \di x^2\wedge \di x^3$. One can observe that, applied to a $k$-vector or $k$-covector, the inverse of the Hodge star is given as
\begin{equation}\label{eq:HodgeInv}
  \hodge^{-1} = (-1)^{k(n-k)} \, \hodge.
\end{equation}
In the special case $n=3$ we also have the following geometric interpretation of the Hodge dual of a $2$-vector: For $\eta \in \Wedge_2 \R^3$, its dual $\hodge \eta$ is an (oriented) normal vector to any two-dimensional hyperplane with orientation $\eta$. Moreover, for $a,b \in \Wedge_1 \R^3 \cong \R^3$ the identities
\begin{equation}\label{eq:CrossvsWedge}
  \hodge(a \times b) = a \wedge b,  \qquad
  \hodge (a \wedge b) = a \times b
\end{equation}
hold, where \enquote{$\times$} denotes the classical vector cross product in $\R^3$. Indeed, for any $v \in\Wedge_1 \R^3$, the triple product $v \cdot (a \times b)$ is equal to the determinant $\det(v,a,b)$ of matrix with columns $v,a,b$, and so
\[
  v \wedge \hodge(a \times b)
  = v \cdot (a \times b) \, \ee_1 \wedge \ee_2 \wedge \ee_3
  = \det(v,a,b) \, \ee_1 \wedge \ee_2 \wedge \ee_3
  = v \wedge (a \wedge b).
\]
Hence, the first identity in~\eqref{eq:CrossvsWedge} follows. The second identity follows by applying $\hodge$ on both sides and using $\hodge^{-1} = \hodge$ (this is~\eqref{eq:HodgeInv} with $n=3$ and $k=1$). Note that no similar representation of a $2$-plane by a normal vector holds in $\R^4$ since the Hodge dual of a $2$-vector in $\R^4$ is again a $2$-vector.

Occasionally, we use the $k$-times \term{wedge product of a linear map}, particularly in the case of the projections $\tbf$ and $\pbf$ defined above. If $S \in \Lcal(\R^n;\R^N)$, we let $\Wedge^k S \in \Lcal(\Wedge^k \R^n; \Wedge^k \R^N)$ be the unique linear map for which
\[
  \Wedge^k S(v_1 \wedge \cdots \wedge v_k) = Sv_1 \wedge \cdots \wedge Sv_k, \qquad v_1, \ldots, v_k \in \R^n;
\]
then we extend by multi-linearity to all of $\Wedge^k S$. For reasons of convenience, we will still write simply $S$ for $\Wedge^k S$.

\subsection{Integral currents} \label{sc:curr}

The theory of currents will provide us with a precise mathematical language to describe dislocations and slip trajectories. For details and proofs of the following statements we refer to~\cite{KrantzParks08book} or the monolithic~\cite{Federer69book}.

Given an open set $U \subset \R^n$ and $k \in \{0,1,\ldots,n\}$, we let $\Dcal^k(U)$ be the space of \term{(smooth) differential $k$-forms} with compact support in $U$, that is, $\Dcal^k(U) := \Crm^\infty_c(U;\Wedge^k \R^n)$. As with vectors and covectors, a $k$-form is called \term{simple} if it takes simple $k$-covector as values everywhere. The \term{exterior differential} of $\omega \in \Dcal^k(U)$ is the $(k+1)$-form $d\omega \in \Dcal^{k+1}(U)$ defined inductively as follows: For a $0$-form $f \in \Dcal^0(U) = \Crm^\infty_c(U;\R)$, we set
\[
  df := \sum_i \frac{\partial f}{\partial x^i} \dd x^i  \quad \in \Dcal^1(U),
\]
where we recall from Section~\ref{sc:algebra} that $\di x^i$ is the $i$th element of the dual canonical basis. Then, for a simple $k$-form which can be expressed as $\omega = f \dd x^{j_1} \wedge \cdots \wedge \di x^{j_k}$ we inductively set $d\omega := (df) \wedge \di x^{j_1} \wedge \cdots \wedge \di x^{j_k}$. For a more general $k$-form, this definition is then extended by linearity.

Just like in the theory of Schwartz distributions, currents are constructed by duality: Indeed, the elements of the dual space $\Dcal_k(U) := \Dcal^k(U)^*$ (with respect to a suitable topology) are called \term{(de Rham) $k$-currents}. This space is equipped with a natural \term{boundary operator}, which for $k$-current $T \in \Dcal^k(\R^n)$, $k \geq 1$, is defined as the $(k-1)$-current $\partial T \in \Dcal_{k-1}(\R^n)$ with
\[
  \dprb{\partial T, \omega} := \dprb{T, d\omega}, \qquad \omega \in \Dcal^{k-1}(\R^n).
\]
For a $0$-current $T$, one formally sets $\partial T := 0$.

We think of $k$-currents as generalized $k$-surfaces. This point of view is particularly pertinent for the following subclass of currents: A (local) Borel measure $T \in \Mcal_\loc(\R^n;\Wedge_k \R^n)$ (i.e., with values in the space $\Wedge_k \R^n$ of $k$-vectors) is called an \term{integer-multiplicity rectifiable $k$-current} ($k \in \N \cup \{0\}$) if it is of the form
\[
  T = m \, \vec{T} \, \Hcal^k \restrict R,
\]
that is, $T$ is an element of the dual space to the space $\Dcal^k(\R^n)$ of $k$-forms via
\[
  \dprb{T,\omega} = \int_R \dprb{\vec{T}(x), \omega(x)} \, m(x) \dd \Hcal^k(x),  \qquad \omega \in \Dcal^k(\R^n),
\]
where
\begin{enumerate}[(i)]
	\item $R \subset \R^n$ is countably $\Hcal^k$-rectifiable with $\Hcal^k(R \cap K) < \infty$ for all compact sets $K \subset \R^n$;
	\item $\vec{T} \colon R \to \Wedge_k \R^n$ is $\Hcal^k$-measurable and for $\Hcal^k$-a.e.\ $x \in R$ the $k$-vector $\vec{T}(x)$ is simple, has unit length ($\abs{\vec{T}(x)} = 1$), and lies in the approximate tangent space $\Tan_x R$ to $R$ at $x$;
	\item $m \colon R \to \N$ is locally $(\Hcal \restrict R)$-integrable.
\end{enumerate}
The $k$-vector field $\vec{T}$ is called the \term{orientation map} of $T$ and $m$ is the \term{multiplicity}. Here, we do not recall precisely the notion of the $k$-dimensional Hausdorff (outer) measure $\Hcal^k$ and (countably) $k$-rectifiable sets (see again~\cite{Federer69book,KrantzParks08book} or~\cite{AmbrosioFuscoPallara00book} for this). Intuitively, one can think of an integer-multiplicity rectifiable $k$-current as a collection of (oriented) $k$-dimensional $\Crm^1$- or Lipschitz-manifolds, which may overlap and thus have a multiplicity other than $1$.

We further have the \term{Radon--Nikod\'{y}m decomposition}
\[
  T = \vec{T} \tv{T}, 
\]
where $\vec{T}$ is the orienting map as above and $\tv{T} = m \, \Hcal^k \restrict R \in \Mcal^+([0,T] \times \R^3)$ is the \term{total variation measure}. The (global) mass of $T$ is
\[
  \Mbf(T) = \tv{T}(\R^n) = \int_R m(x) \dd \Hcal^k(x).
\]

The following classes of \term{integral $k$-currents} ($k \in \N \cup \{0\}$) are central to our rigorous theory:
\begin{align*}
  \Irm_k(\R^n) &:= \setb{ \text{$T$ integer-multiplicity rectifiable $k$-current} }{ \Mbf(T) + \Mbf(\partial T) < \infty }, \\
  \Irm_k(\cl{\Omega}) &:= \setb{ T \in \Irm_k(\R^n) }{ \supp T \subset \cl{\Omega} }.
\end{align*}
By the so-called boundary rectifiability theorem, see~\cite[4.2.16]{Federer69book} or~\cite[Theorem~7.9.3]{KrantzParks08book}, for $T \in \Irm_k(\cl{\Omega})$, also $\partial T \in \Irm_{k-1}(\cl{\Omega})$.

We also briefly recall the theory of pushforwards of integral currents. Let $\theta \colon \cl{\Omega} \to \R^n$ be smooth and let $T = m \, \vec{T} \Hcal^k \restrict R \in \Irm_k(\cl{\Omega})$. The \term{(geometric) pushforward} $\theta_* T$ (often denoted by \enquote{$\theta_\# T$} in the geometric measure theory literature) is defined via
\[
  \dprb{\theta_* T,\omega} := \dprb{T, \theta^*\omega}, \qquad \omega \in \Dcal^k(\R^n),
\]
where $\theta^*\omega$ denotes the pullback of the $k$-form $\omega$. If $\theta|_{\supp T}$ is \term{proper}, i.e., $\theta^{-1}(K) \cap \supp T$ is compact for every compact $K \subset \R^n$, then $\theta_* T \in \Irm_k(\cl{\theta(\Omega)})$, see, for instance,~\cite[(3) on p.197]{KrantzParks08book}. If we denote the approximate derivative of $\theta$ (which is defined almost everywhere) with respect to the $k$-rectifiable set $R$ carrying $T$ by $D^R \theta$ (i.e. $D^R \theta$ is the restriction of $D\theta$ to $\Trm_x R$), then
\[
  \dprb{\theta_* T, \omega} = \int \dprb{D^R\theta(\vec{T}(x)), \omega(\theta(x))} \dd \tv{T}(x),  \qquad \omega \in \Dcal^k(\R^n).
\]
It also holds that
\begin{equation} \label{eq:pushforward_bdry}
  \partial (\theta_* T) = \theta_*(\partial T).
\end{equation}

Finally, the slicing theory of integral currents (see~\cite[Section~7.6]{KrantzParks08book} or~\cite[Section~4.3]{Federer69book}) allows one to define the \enquote{restriction} of a given integral current $S = m \, \vec{S} \, \Hcal^{k+1} \restrict R \in \Irm_{k+1}(\cl{\Omega})$ to a level set of a Lipschitz map $f \colon \R^n \to \R$ (for us, this will always be the temporal projection $\tbf(t,x) := t$) as follows: Set $R|_t := f^{-1}(\{t\}) \cap R$. Then, $R|_t$ is (countably) $\Hcal^k$-rectifiable for almost every $t \in \R$ and for $\Hcal^{k+1}$-almost every $z \in R|_t$ (if $f = \tbf$, then $z = (t,x)$), the approximate tangent spaces $\Trm_z R$ and $\Trm_z R|_t$, as well as the approximate gradient $\nabla^R f(z)$ (the projection of $\nabla f(z)$ onto $\Trm_z R$) exist. Furhermore,
\[
  \Trm_z R = \spn \bigl\{ \Trm_z R|_t, \xi(z) \bigr\},  \qquad
  \xi(z) := \frac{\nabla^R f(z)}{\abs{\nabla^R f(z)}} \perp \Trm_z R|_t.
\]
Also, $\xi(z)$ is simple and has unit length. If we set
\[
  m_+(z) := \begin{cases}
    m(z) &\text{if $\nabla^R f(z) \neq 0$,} \\
    0    &\text{otherwise,}  
  \end{cases}
  \qquad\qquad
  \xi^*(z) := \frac{D^R f(z)}{\abs{D^R f(z)}},
\]
where $D^R f(z)$ is the restriction of the differential $Df(z)$ to $\Trm_z R$,and
\[
  \vec{S}|_t(z) := \vec{S}(z) \restrict \xi^*(z) 
  \in \Wedge_k \Trm_z R|_t
  \subset \Wedge_k \Trm_z R,
\]
then we may define the \term{slice} of $S$ with respect to $f$ at $t$ via
\[
  S|_t := m_+ \, \vec{S}|_t \, \Hcal^k \restrict R|_t.
\]
It can be shown that $S|_t$ is an integral $k$-current satisfying in particular the following properties:
\begin{enumerate}[(i)]
\item \emph{The coarea formula for slices:}
\begin{equation} \label{eq:coarea_slice}
  \int_R g \, \abs{\nabla^R f} \dd \Hcal^{k+1} = \int \int_{R|_t} g \dd \Hcal^k \dd t
\end{equation}
holds for all $g \colon R \to \R^N$ that are $\Hcal^{k+1}$-measurable and integrable on $R$.
\item \emph{The mass decomposition:}
\[
  \int \Mbf(S|_t) \dd t = \int_R \abs{\nabla^R f} \dd \tv{S}.
\]
\item \emph{The cylinder formula:}
\[
  S|_t = \partial(S \restrict \{f<t\}) - (\partial S) \restrict \{f<t\}.
\]
\item \emph{The boundary formula:}
\[
  \partial (S|_t) = - (\partial S)|_t.
\]
\end{enumerate}

\subsection{Description of dislocations and slip trajectories via integral currents} \label{sc:geom}

Our modeling of dislocations is made rigorous as follows: All dislocation lines with structural Burgers vector $b \in \Bcal$ that are contained in our material body at time $t \in [0,T]$, are collected in a $1$-dimensional integral current $T^b(t)$ in $\cl{\Omega}$ (see~\cite{ContiGarroniMassaccesi15,ContiGarroniOrtiz15,ScalaVanGoethem19} for similar approaches). This current is boundaryless in $\Omega$, i.e.,
\[
  \partial T^b(t) \restrict \Omega = 0
\]
because dislocation lines are always closed loops inside the specimen $\Omega$. Moreover, since one may flip the sign of a Burgers vector when at the same time also reversing all dislocation line directions, the symmetry relation
\[
  T^{-b}(t) = -T^b(t)
\]
needs to hold for the family $(T^b(t))_{b \in \Bcal}$ and (almost) every $t \in [0,T]$.

The associated \term{slip trajectories} likewise can be expressed as $2$-dimensional integral currents $S^b$ (for the Burgers vector $b \in \Bcal$) in the space-time cylinder $[0,T] \times \Omega$ with the property that
\begin{equation} \label{eq:Sb_bdry_ax}
  \partial S^b \restrict ((0,T) \times \Omega) = 0.
\end{equation}
Moreover, the symmetry relation
\[
  S^{-b} = -S^b
\]
needs to hold for the family $(S^b)_{b \in \Bcal}$, just like it did for the dislocations themselves. In this description via slip trajectories, the \term{dislocation system} at time $t$ is given by
\begin{equation} \label{eq:Tb}
  T^b(t) := \pbf_*(S^b|_t),  \qquad b \in \Bcal,
\end{equation}
i.e., the pushforward under the spatial projection $\pbf(t,x) := x$ of the slice $S^b|_t$ of $S^b$ at time $t$ (that is, with respect to the temporal projection $\tbf(t,x) := t$). The pushforward here moves the slice from $\{t\} \times \Omega$ to $\Omega$. The slicing theory of integral currents recalled in the previous section entails that $T^b(t)$ is a $1$-dimensional integral current in $\Omega$. Further,~\eqref{eq:Sb_bdry_ax} implies $\partial T^b(t) \restrict \Omega = 0$ and $T^{-b}(t) = -T^b(t)$ for almost every $t \in (0,T)$ and all $b \in \Bcal$.

The total traversed \term{slip surface} from $T^b(s)$ to $T^b(t)$ is the integral $2$-current in $\R^3$ given by
\[
  S^b|_s^t := \pbf_* \bigl[ S^b \restrict ([s,t] \times \Omega) \bigr],
\]
that is, the pushforward under the spatial projection of the restriction of $S^b$ to the time interval $[s,t]$. However, $S^b|_s^t$ does not contain a \enquote{time index}, i.e., a time coordinate, which is needed to define the plastic flow (see Section~\ref{sc:plastflow}). Furthermore, areas that are traversed several times may lead to cancellations in the slip surface $S^b|_s^t$, but not in the slip trajectory $S^b$ itself. These are the fundamental reasons why we work with the space-time slip trajectories and not the slip surfaces in this work.

We further assume that there are no \enquote{horizontal parts} in $S^b$, that is,
\begin{equation} \label{eq:no_horz}
  \tv{S^b}(\{t\} \times \Omega) = 0  \qquad\text{for a.e.\ $t \in [0,T]$.}
\end{equation}
The reason for this is that over such horizontal parts we again lack the \enquote{time index} we need to define the plastic evolution (a horizontal piece is essentially just a slip surface that gets traversed instantaneously).

The companion work~\cite{Rindler21a?} investigates the notion of space-time trajectories, and in particular the associated notion of \emph{variation} (the simplest form of a dissipation along a slip trajectory), in more detail.

\subsection{Dislocation velocity} \label{sc:slipvel}

The dislocation velocity is a natural quantity when considering space-time integral currents: Let us first recall from the slicing theory of integral currents that for $\Lcal^1$-almost every $t \in [0,T]$ and $\tv{S^b|_t}$-almost every $(t,x)$ the approximate tangent spaces $\Trm_{(t,x)} S^b$, $\Trm_{(t,x)} S^b|_t$ as well as the approximate gradient $\nabla^{S^b} \tbf(t,x)$ to $\tbf$ at $(t,x)$ (that is, the orthogonal projection of $\nabla \tbf(t,x)$ onto $\Trm_{(t,x)} S^b$) exist (see, e.g.,~\cite[Lemma~7.6.1~(2)]{KrantzParks08book}). Here we also use \enquote{$S^b$} in place of its carrier set, i.e., in place of $R^b$ if $S^b = m^b \, \vec{S}^b \, \Hcal^2 \restrict R^b$. It also holds that
\[
  \Trm_{(t,x)} S^b = \spn \bigl\{ \Trm_{(t,x)} S^b|_t, \xi^b(t,x) \bigr\},  \qquad
  \xi^b(t,x) := \frac{\nabla^{S^b} \tbf(t,x)}{\abs{\nabla^{S^b} \tbf(t,x)}} \perp \Trm_{(t,x)} S^b|_t.
\]
Thus,
\[
  \vec{S}^b|_t = \vec{S}^b \restrict \xi^b,  \qquad
  \vec{S}^b = \xi^b \wedge \vec{S}^b|_t,
\]
where the second equality is a consequence of the general relation $(\xi \wedge \tau) \restrict \xi^* = \tau$ for any $\tau \in \Wedge_k \R^{1+d}$ with $\tau \restrict \xi^* = 0$ (see~\cite[1.5.3]{Federer69book}).

In the following we fix $t,x$ as above. We calculate (note $\abs{\xi^b} = 1$)
\[
  \abs{\nabla^{S^b} \tbf}
  = \abs{(\ee_0 \cdot \xi^b)\xi^b}
  = \abs{\xi^b \cdot \ee_0}
  = \abs{(\xi^b \cdot \ee_0)\ee_0}
  = \abs{\tbf(\xi^b)}
\]
and (recall $\pbf(\xi^b) \perp \vec{S}^b|_t$)
\[
  \abs{\pbf(\vec{S}^b)}
  = \abs{\pbf(\xi^b) \wedge \vec{S}^b|_t}
  = \abs{\pbf(\xi^b)}.
\]
Since $\abs{\tbf(\xi^b)}^2 + \abs{\pbf(\xi^b)}^2 = 1$, we obtain in particular the \enquote{Pythagorean} identity
\begin{equation} \label{eq:decomp}
  \abs{\nabla^{S^b} \tbf}^2 + \abs{\pbf(\vec{S}^b)}^2 = 1  \qquad \text{$\tv{S}$-a.e.}
\end{equation}
Thus, our definition of the dislocation velocity,
\begin{equation} \label{eq:slipvel_geo}
  \frac{\DD}{\DD t} S^b(t,x)
  := \frac{\pbf(\xi^b(t,x))}{\abs{\tbf(\xi^b(t,x))}}
  = \frac{\pbf(\xi^b(t,x))}{\abs{\nabla^{S^b} \tbf(t,x)}}
  = \frac{\pbf(\xi^b(t,x))}{\sqrt{1-\abs{\pbf(\vec{S}^b)(t,x)}^2}}
\end{equation}
expresses indeed the velocity of dislocation motion.

\subsection{Infinitesimal plastic shear} \label{sc:slipderivation}

In this section we provide an alternative, more precise, derivation of the formula for the infinitesimal plastic shear~\eqref{eq:DtR}, which takes into account the transformation of planes between the reference configuration and the structural space.

Consider an open and bounded mesoscopic reference domain $U \subset \Omega$, where \enquote{mesoscopic} means that atomistic effects can be neglected, but the deformation is nearly constant in $U$. Assume without loss of generality that $0 \in U$ and that $U$ is partitioned into two parts by an oriented hyperplane $H$. We describe the hyperplane $H$ by a $2$-vector $\nu \in \Wedge_2 \R^3$ and denote its associated (oriented) normal vector by $N = \hodge \nu$, where $\hodge$ is the Hodge star.

Fix a (structural) Burgers vector $b \in \Bcal$ and assume that we are investigating a slip trajectory $S^b$ such that the dislocations at any time $t\in[0,T]$, i.e., $T^b(t) = \pbf_*(S^b|_t)$, lie entirely in the hyperplane $H$. Locally, this situation approximately describes the general picture with $H$ being (the spatial restriction of) the approximate tangent space to $S^b$ at a point $(t_0,x_0) \in (0,T) \times \Omega$. Up to time $t$, the dislocations in $S^b$ traverse the restricted slip trajectory $S^b \restrict ([0,t] \times U)$. At time $t$, the dislocations are at the leading edge of $S^b$, i.e., $T^b(t) = \partial [S^b \restrict ([0,t] \times U)] + T_0^b$, where $T_0^b$ are the dislocations at the beginning of the evolution represented by $S^b$.

Assume first that $H$ is precisely the lattice plane dual to the bond $q_i$ (the $i$th scaffold vector), which is the plane with orientation $q_j \wedge q_k$ such that $ijk$ is an even permutation of $123$, meaning that $\nu = q_j \wedge q_k$ (e.g., if we are considering the bond $q_1$, then the dual plane is $\nu = q_2 \wedge q_3$). As the dislocation moves, the $i$th bond is translated by the opposite of the structural Burgers vector $b$. Thus, since the $q_i$ are expressed in referential coordinates, $q_i$ is mapped to $q_i - Qb$. For any other (referential) vector $w$ we posit that only the part of $w$ in direction $q_i$ gets translated by the corresponding fraction of $Qb$. Furthermore, if $H$ is not aligned with a lattice plane, then we assume that $H$ is linearly decomposed into a sum of its components in the lattice planes $q_j \wedge q_k$ for $j \neq k$ (the decomposition is with respect to the linear structure of $\Wedge_2 \R^3$) and the above translation happens separately for all these lattice planes. Summing up, a (referential) vector $w$ is transformed by the dislocations into
\begin{equation} \label{eq:v_transf}
  w' := w - (Qb)\alpha(\nu,w),
\end{equation}
where $\alpha = \alpha_{t,x} \colon \Wedge_2 \R^3 \times \R^3 \to \R$ has the following properties:
\begin{equation} \label{eq:alpha}
  \left\{ \begin{aligned}
    &\text{$\alpha(\frarg,\frarg)$ is bilinear;}\\
    &\text{$\alpha(q_j \wedge q_k, q_\ell) = \delta_{\ell i}$ for $ijk$ an even permutation of $123$.}
  \end{aligned} \right.
\end{equation}
We remark that this requirement is correct even if the lengths of scaffold vectors $q_i$ have changed. For instance, if $q_1 = 2\ee_1$, $q_2 = \ee_2$, $q_3 = \ee_3/2$ (so that $\det Q = \det(q_1,q_2,q_3) = 1$) then the plane $\ee_1 \wedge \ee_2$ represents \enquote{half} a lattice plane (because the lattice has changed) and so there should only be half the corresponding transformation applied.

We collect the scaffold vectors $q_1, q_2, q_3$ at $(t,x)$ into the matrix $Q := (q_1,q_2,q_3)$ (as columns), which by the assumed plastic incompressibility (see Section~\ref{sc:restrict}) satisfies $\det Q = 1$. Then, we claim that the (unique) map $\alpha$ that satisfies the properties given in~\eqref{eq:alpha} is
\begin{equation} \label{eq:hodgeQ}
  \alpha(\xi,w) = [Q^{-T} \hodge Q^{-1} \xi]^T w  = (\hodge \xi)^T w,  \qquad \xi \in \Wedge_2 \R^3, \; w \in \R^3.
\end{equation}
Here, the expression $Q^{-T} \hodge Q^{-1} \xi$ is to be bracketed from the right, i.e., to be understood as $Q^{-T} (\hodge(Q^{-1} \xi))$. The first equality follows simply because this expression for $\alpha$ (uniquely) satisfies the conditions in~\eqref{eq:alpha}, as can be easily checked. To see the equality between the second and third expression in~\eqref{eq:hodgeQ}, we argue as follows: First observe that any simple $3$-vector $\zeta = z_1 \wedge z_2 \wedge z_3$ in $\R^3$ can be written as $\det(z_1,z_2,z_2) \, \Erm$ with $\Erm := \ee_1 \wedge \ee_2 \wedge \ee_3$, and hence
\[
  Q^T \zeta = (Q^T z_1) \wedge (Q^T z_2) \wedge (Q^T z_3)
  = \det(Q^T z_1,Q^T z_2,Q^T z_3) \, \Erm
  = \det Q \cdot \det(z_1,z_2,z_3) \, \Erm
  = \zeta
\]
since $\det Q = 1$ by the assumed plastic incompressibility. Thus, using the definition of the Hodge star, for any $\eta \in \Wedge_2 \R^3$ we have
\begin{align*}
  \eta \wedge [Q^{-T} \hodge Q^{-1} \xi]
  &= (Q^T \eta) \wedge [\hodge Q^{-1} \xi] \\
  &= (Q^T \eta, Q^{-1} \xi) \, \Erm \\
  &= (\eta, \xi) \, \Erm \\
  &= \eta \wedge (\hodge \xi),
\end{align*}
and so $Q^{-T} \hodge Q^{-1} \xi = \hodge \xi$, completing the proof of the second equality in~\eqref{eq:hodgeQ}.

If we apply~\eqref{eq:v_transf} with the $\alpha$ identified above to $w = q_1,q_2,q_3$ (which form a basis), then we obtain that the scaffold $Q$ is transformed as
\begin{equation} \label{eq:Qprime}
  Q \mapsto Q' 
  := Q - (Qb)(\hodge \nu)^T Q
  = \bigl( \Id - (Qb) \otimes (\hodge \nu) \bigr) Q
\end{equation}
at time $t$ such that $(t,x) \in \supp T^b(t)$. Also considering the multiplicity $\phi^b(t) = m^b(t) = \tv{T^b(t)}$ of the moving dislocations at time $t$ and recalling that $\hodge \nu = N$ (the weighted normal to $H$), we obtain~\eqref{eq:DtR}.

\subsection{Plastic flow} \label{sc:plastflow_rigorous}

In this section we provide additional details concerning the derivation of the plastic flow equation~\eqref{eq:plast_flow}. We start from~\eqref{eq:DtR}, which means that from time $t$ to $t+\delta$ we progress from $Q$ to $Q'$ given as
\[
  Q' := \bigl( \Id - (Qb) \otimes \bigl[\hodge \pbf(S^b_\eta)((t,t+\delta) \times \Omega)\bigr] \bigr) Q.
\]
Here we have set
\[
  \pbf(S^b_\eta) := \pbf(\vec{S}^b_\eta) \, \tv{S^b_\eta} \in \Mcal([0,T] \times \Omega;\Wedge_2 \R^3),
\]
and we have used that locally around $(t,x)$ we slip over the plane $H$ with normal $N = \hodge\nu = \hodge \pbf(S^b_\eta(t,x))$ (we have already smeared out $S^b$ via the dislocation line profile $\eta$).

In order to rewrite the above relation as a differential equation, we now need to identify the \emph{density} of the measure on the right with respect to Lebesgue measure. We first observe $\abs{\pbf(\vec{S}^b(t,x))} < 1$ for $\tv{S^b}$-almost every $(t,x)$. Otherwise, $\pbf(\vec{S}(t,x)) = \vec{S}(t,x)$ and there would exist a vertical piece in $\tv{S^b}$, i.e., there would be a $t_0 \in [0,T]$ such that $\tv{S^b}(\{t_0\} \times \R^3) > 0$, contradicting~\eqref{eq:no_horz}. So, by~\eqref{eq:decomp},
\[
  \nabla^{S^b} \tbf(t,x) \neq 0  \qquad\text{for $\tv{S^b}$-almost every $(t,x)$.}
\]
From the coarea formula for slices, see~\eqref{eq:coarea_slice}, we then get for any $\omega \in \Dcal^2(\R^{1+3})$ that
\begin{align}
  \dprb{\pbf(S^b_\eta),\omega} &= \int \dprb{\pbf(\vec{S}^b(t,x)), [\eta \conv \omega(t,\frarg)](x)} \dd \tv{S^b}(t,x)  \notag\\
  &= \int_0^1 \int \dprBB{\frac{\pbf(\vec{S}^b(t,x)) }{\abs{\nabla^{S^b} \tbf(t,x)}}, [\eta \conv \omega(t,\frarg)](x)} \dd \tv{S^b|_t}(x) \dd t  \notag\\
  &= \int_0^1 \int \dprBB{\eta \conv \biggl[ \frac{\pbf(\vec{S}^b(t,\frarg)) \tv{S^b|_t}}{\abs{\nabla^{S^b} \tbf(t,\frarg)}} \biggr], \omega} \dd x \dd t.  \label{eq:Sb_transform}
\end{align}
The expression on the left in the last duality bracket is thus the density of the measure $\pbf(S^b_\eta)$, which we call the $2$-vector version of the \term{geometric slip rate} $\gamma^b = \gamma^b(t,x) \in \Wedge_2 \R^3$. Using also the definition of the dislocation velocity in~\eqref{eq:slipvel_geo},
\[
  \gamma^b(t,\frarg)
  = \eta \conv \biggl[ \frac{\DD}{\DD t} S^b(t,\frarg) \wedge \pbf(\vec{S}^b|_t) \, \tv{S^b|_t} \biggr]
  = \eta \conv \biggl[ \frac{\DD}{\DD t} S^b(t,\frarg) \wedge T^b(t) \biggr],
\]
which is~\eqref{eq:gammab}.

The above arguments, in particular~\eqref{eq:Qprime}, together with the definition $g^b := \hodge\gamma^b$ of the \term{normal geometric slip rate} (called only \enquote{geometric slip rate} in Section~\ref{sc:plastflow}) entail that in the time interval $[t,t+\delta]$ ($\delta > 0$) the scaffold $Q$ changes to approximately
\[
  Q(t+\delta) \approx \biggl( \Id - (Q(t)b) \otimes \int_t^{t+\delta} g^b(\tau) \dd \tau \biggr) Q(t).
\]
Thus,
\[
  - Q(t)^{-1} \frac{Q(t+\delta) - Q(t)}{\delta} Q(t)^{-1} = b \otimes \frac{1}{\delta} \int_t^{t+\delta} g^b(\tau) \dd \tau
\]
Letting $\delta \to 0$, we arrive at
\[
  \dot{P} = -Q^{-1}\dot{Q}Q^{-1} = b \otimes g^b(t).
\]
It remains to integrate against the Burgers measure $\kappa$ to take into account the slip for all Burgers vectors $b$ to obtain the plastic flow equation~\eqref{eq:plast_flow} and the definition of the total plastic drift in~\eqref{eq:D}.

\subsection{Proof the consistency condition} \label{sc:consistproof}

To prove the consistency condition~\eqref{eq:consist} at $t > 0$, we assume that this equation holds at $t = 0$ (as an initial condition) and then argue as follows: Recall the plastic flow equation~\eqref{eq:plast_flow} and~\eqref{eq:D}, which read as
\[
  \dot{P} = \int b \otimes  g^b \dd \kappa(b).
\]
We integrate this in time from $0$ to $t$ and apply the row-wise $\curl = \nabla \times$ (in space) to both sides, to obtain
\begin{equation} \label{eq:consist_1}
  \curl P(t) - \curl P(0)
  = \int b \otimes \biggl( \curl \int_0^t g^b(\tau) \dd \tau \biggr) \dd \kappa(b).
\end{equation}
To evaluate the right-hand side, we use that $g^b = \hodge\gamma^b$ and observe
\[
  \int_0^t g^b(\tau) \dd \tau
  = \int_0^t \hodge \gamma^b(\tau) \dd \tau
  = \hodge \int_0^t \gamma^b(\tau) \dd \tau
  = \hodge \pbf_*\bigl( S^b_\eta \restrict ((0,t) \times \Omega) \bigr)
\]
by similar arguments as in~\eqref{eq:Sb_transform}. Using that \enquote{$\curl \circ \, \hodge = \partial$}, which will be shown below, as well as~\eqref{eq:pushforward_bdry},~\eqref{eq:Sb_bdry_ax}, and $T^b_\eta(t) = \pbf_*(S^b_\eta|_t)$,  we then obtain
\begin{align*}
  \curl \int_0^t g^b(\tau) \dd \tau
  &= \partial \bigl[ \pbf_*\bigl( S^b_\eta \restrict ((0,t) \times \Omega) \bigr) \bigr] \\
  &= \pbf_* \bigl[ \partial \bigl( S^b_\eta \restrict ((0,t) \times \Omega) \bigr) \bigr] \\
  &= \pbf_* \bigl[ S^b_\eta|_t - S^b_\eta|_0 \bigr] \\
  &= T^b_\eta(t) - T^b_\eta(0).
\end{align*}
Note that we can only do this at $t$ for which the slice $S^b_\eta|_t$ is defined, but this is the case almost everywhere in $(0,T)$. Plugging the last formula into~\eqref{eq:consist_1},
\[
  \curl P(t) - \curl P(0)
  = \int b \otimes T^b_\eta(t) \dd \kappa(b) - \int b \otimes T^b_\eta(0) \dd \kappa(b).
\]
Finally, using the consistency condition~\eqref{eq:consist} at $t = 0$, we arrive at~\eqref{eq:consist} for every $t$ where the slice is defined.

It remains to show the identity \enquote{$\curl \circ \, \hodge = \partial$}. To do so, we recall that the Hodge dual for a $k$-covector $\alpha$ satisfies
\[
  \hodge \alpha = (\hodge \alpha^\sharp)^\flat
\]
with the musical isomorphisms $\flat \colon \Wedge_k V \to \Wedge^k V$ (\enquote{lowering an index}) and $\sharp \colon \Wedge^k V \to \Wedge_k V$ (\enquote{raising an index}), which are determined by
\[
  \dprb{\xi,\alpha} = \sprb{\xi^\flat,\alpha} = \sprb{\xi,\alpha^\sharp},  \qquad \xi \in \Wedge_k V, \; \alpha \in \Wedge^k V.
\]
Then, let $S \in \Irm_2(\R^3)$ and compute, using the symmetry of the $\curl$ operator, the formula $(\curl \omega^\sharp)^\flat = \hodge d\omega$ for any $\omega \in \Dcal^1(\R^3)$, and the duality rule $\dpr{\hodge \xi, \hodge \alpha} = \dpr{\xi,\alpha}$, as follows:
\[
  \dprb{\curl(\hodge S), \omega}
  = \dprb{\hodge S, (\curl \omega^\sharp)^\flat}
  = \dprb{\hodge S, \hodge d\omega}
  = \dprb{S, d\omega}
  = \dprb{\partial S, \omega},
\]
which proves the claim.

\appendix

\section{Comparison to geometric and constitutive paradigms} \label{ax:comparison}

In this appendix we explain the relationship of our kinematic modeling with the \enquote{geometric paradigm} by Kondo~\cite{Kondo55}, Nye~\cite{Nye53}, Bilby~\cite{BilbyBulloughSmith55} and Kr\"{oner}~\cite{Kroner01,Kroner60,Kroner61} as well as the \enquote{constitutive paradigm} by Noll~\cite{Noll58} and Wang~\cite{Wang67}. The general idea of the first of these paradigms is that the microstructure of a material is encoded in additional geometric data, usually another locally flat metric connection, which is attached to the body (manifold). The second of these paradigms starts from an energy density \enquote{archetype}, which is implanted by means of a given \enquote{implant} frame field. The two paradigms are compatible and complementary if certain conditions on the topology and the symmetry of the energy density hold. We refer to~\cite{EpsteinKupfermanMaor20} for a recent survey as well as many historical remarks and references.

The model developed in the present work sits in between both approaches: A dislocation system $(T^b)_{b \in \Bcal}$ can be considered additional \enquote{geometric data} on the manifold, so falls into the geometric paradigm, which is usually formulated for \enquote{continuous} distributions of dislocations rather than the discrete dislocations that we consider here. In fact, as we will see below, dislocations are precisely dual to the torsion of the additional connection on the body (the torsion being the manifestation of dislocations in the geometric paradigm). On the other hand, the scaffold frame $Q$ turns out to be the implant map of the constitutive paradigm. It is precisely our consistency relation~\eqref{eq:consist}, which is preserved along the plastic flow, that allows us to use both paradigms for our description of the dislocation motion and plastic dynamics.


\subsection{Torsion and Burgers vector}

In the geometric paradigm (we follow~\cite{EpsteinKupfermanMaor20} for terminology and notation) one starts with a smooth $3$-dimensional Riemannian \term{body manifold} $(M,g)$ with boundary together with an additional affine connection $\nabla$ on $M$. We also require that $\nabla$ is metrically consistent, i.e., the $\nabla$-parallel transport is an isometry (with respect to $g$), and that $\nabla$ is locally flat, i.e., the associated Riemannian curvature tensor vanishes. The triple $(M,g,\nabla)$ is called a \term{Weitzenb\"{o}ck manifold}. Let $\{q_i|_x\}_i$ be a $\nabla$-parallel frame for $M$ and $\{\vartheta^i|_x\}_i$ its dual coframe (i.e., $\dpr{q_i|_x,\vartheta^j|_x} = \delta_{ij}$). We know that locally such a frame exists since the $\nabla$-parallel transport is locally path-independent by the flatness of $\nabla$. Let us assume that this frame exists \emph{globally}, meaning that the $\nabla$-parallel transport is globally path-independent (thus constituting a \enquote{distant parallelism}). This is chiefly a topological restriction on $M$; see~\cite{EpsteinKupfermanMaor20} for more on this point.

Alternatively, one can start from a (global or local) frame $\{q_i|_x\}_i$ and define a locally flat connection $\nabla$, for which $\{q_i|_x\}_i$ is $\nabla$-parallel. In both cases, the frame $\{q_i|_x\}_i$ takes the same role as the scaffold vectors defined in Section~\ref{sc:kinematics}, that is (a spanning subset of) the lattice vectors of the crystal relative to the referential positions of the atoms. We thus refer to it as the \term{scaffold frame}.

The crucial point is that the scaffold frame $\{q_i|_x\}_i$ may have \emph{torsion}. That is, the $\Trm M$-valued \term{torsion $2$-form} of $\nabla$,
\[
  \tau(X,Y) := \nabla_X Y - \nabla_Y X - [X,Y]
\]
is in general non-vanishing. For the locally flat connection $\nabla$ we have
\[
  \tau = q_i \otimes d\vartheta^i.
\]
Indeed,
\[
  d\vartheta^i(q_j,q_k)
  = q_j(\vartheta^i(q_k)) - q_k(\vartheta^i(q_j)) - \vartheta^i([q_j,q_k])
  = -\vartheta^i([q_j,q_k])
  = \vartheta^i(\tau(q_j,q_k))
\]
since $\tau(q_j,q_k) = -[q_j,q_k]$ (note that we here consider $q_j$ to be identified with its tangential directional derivative). Thus, $d\vartheta^i = \vartheta^i \circ \tau$ and hence $\tau = q_i \otimes d\vartheta^i$ (where we have now considered $\tau$ as a section of the tensor bundle $\Trm M \otimes \Wedge^2 \Trm M$).

The lattice displacement experienced by an observer traveling along a loop $\gamma \colon [0,1] \to M$ is measured by the \term{Burgers vector} (the \enquote{sum of the tangents} along $\gamma$)
\[
  b_\gamma := \int_0^1 \Pi_{\gamma(t)}^{\gamma(0)} \dot\gamma(t) \dd t \in \Trm_{\gamma(0)} M,
\]
where we have denoted by $\Pi_{\gamma(t)}^{\gamma(0)}$ the $\{q_i|_x\}_i$-parallel transport from $\Trm_{\gamma(t)} M$ to $\Trm_{\gamma(0)} M$. If $b_\gamma$ is non-zero, then $\gamma$ encloses a dislocation line.

\subsection{Dislocation lines via dualization}

Another way to detect the presence of dislocations, dual to the one just presented, is the following: Let $D \subset M$ be a two-dimensional embedded submanifold of $M$ with boundary, e.g.\ a (distorted) disk. The integral
\[
  \int_{\partial D} \vartheta^i = \int_D d\vartheta^i
\]
is known as the \emph{stacking fault} of the lattice planes defined by the $1$-form $\vartheta^i$. Indeed, one may think of the form $\vartheta^i$ in the Ehresmann sense, where $\vartheta^i|_x$ is identified with the $2$-dimensional subspace $\ker \vartheta^i_x$ in $\Trm_x M$. The coframe $\{\vartheta^i|_x\}_i$ thus specifies the lattice planes and the (first) integral measures the signed sum of planes traversed by $\partial D$. The identity between the first and second integral is precisely Stokes's theorem.

We can relate this dual description to the Burgers vector by considering a family $D_\eps$ of (distorted) disks shrinking to a point $x \in M$ and with orientations converging to $\xi \in \Wedge_2 \Trm_x M$ (we argue somewhat heuristically here and in the following to not overburden the explanation with technicalities). Then,
\[
  \frac{\di}{\di \eps}\bigg|_{\eps = 0} b_{\partial D_\eps} = \tau_x(\xi) = q_i|_x \; d\vartheta^i|_x(\xi).
\]
Here, $\partial D_\eps$ denotes the boundary curve traversed in the induced boundary orientation. 

In order to move to the more familiar description of dislocations as loops or, more generally, boundaryless $1$-currents, we need to dualize the above expression for the torsion (see~\cite{EpsteinSegev14} for similar arguments). For this, we define $2$-currents $S_i$ as follows: For a (smooth) $2$-form $\beta$, we let
\[
  \dprb{S_i, \beta} := \int_M \vartheta^i \wedge \beta.
\]
Let us remark that if $M$ is a subset of Euclidean space $\R^3$, then the right-hand side is more concretely written as $\int_M \dpr{\Erm, \vartheta^i \wedge \beta} \dd x$ with the canonical orienting $3$-vector $\Erm := \ee_1 \wedge \ee_2 \wedge \ee_3$ in $\R^3$.

We have
\[
  \partial S_i = T_i
\]
with $T_i$ a $1$-current that is given for a compactly supported smooth $1$-form $\omega$ as
\[
  \dprb{T_i, \omega} := \int_M d\vartheta^i \wedge \omega.
\]
Indeed, using the formula
\[
  \int_M d\vartheta \wedge \omega - \vartheta \wedge d\omega = \int_M d(\vartheta \wedge \omega) = 0,
\]
we obtain
\[
  \dprb{\partial S_i, \omega} = \dprb{S_i, d\omega}
  = \int_M \vartheta^i \wedge d\omega
  = \int_M d\vartheta^i \wedge \omega
  = \dprb{T_i, \omega}.
\]
In particular, we observe (in $M$)
\[
  \partial T_i = \partial \partial S_i = 0.
\]
This dualization allows us to generalize the theory to non-smooth situations via the Federer--Fleming theory of integral and normal currents~\cite{Federer69book,KrantzParks08book}, which is, for instance employed in the works~\cite{ContiGarroniMassaccesi15,ContiGarroniOrtiz15,ScalaVanGoethem19,KupfermanOlami20}.

\subsection{Torsion induced by discrete loops}

Assume now that we are working in three-dimensional Euclidean space $M = \R^3$ (with the canonical metric) and that, formally, the torsion tensor $\tau$ can be written as
\[
  \tau = b \otimes \xi^\flat \; \Hcal^1 \restrict \im \gamma,  \qquad
  \xi := \hodge \frac{\dot\gamma}{\abs{\dot{\gamma}}},
\]
where $\gamma$ is a (smooth or rectifiable) loop and $\Hcal^1 \restrict \im \gamma$ denotes the $1$-dimensional (Hausdorff) measure on the image of $\gamma$, taking into account the multiplicity on overlapping parts of the loop. Note that $\xi$ is the $2$-vector such that $\dot\gamma/\abs{\dot{\gamma}}$ is its oriented normal and $\xi^\flat \in \R^3$ is the $2$-covector dual to $\xi$, so that $\dpr{\xi,\xi^\flat} = 1$. The Burgers vectors $b \in \R^3$ should be seen as fixed along the dislocation only when expressed in the structural frame. However, here we look at a fixed time, and hence there is no problem in leaving the precise nature of $b$ undefined.

Let $\omega$ be a smooth $1$-form and compute (somewhat formally)
\begin{align*}
  \sum_i q_i \, \dprb{T_i,\omega}
  = \int \sum_i q_i \, d \vartheta^i \wedge \omega
  = \int \tau \wedge \omega
  = \int \dprb{\Erm \restrict \tau, \omega} \dd x
  = b \int_{\im \gamma} \dprBB{\frac{\dot\gamma}{\abs{\dot{\gamma}}},\omega} \dd \Hcal^1
\end{align*}
since $\Erm \restrict \xi^b = \hodge \xi = \dot\gamma/\abs{\dot{\gamma}}$. To see this identity, write $\xi = v_1 \wedge v_2$ with $v_1,v_2$ orthonormal. Let $u \in \R^3$ be chosen such that $\Erm = v_1 \wedge v_2 \wedge u$, that is, $u$ extends $\{v_1, v_2\}$ to a positively-oriented orthonormal basis of $\R^3$. It is an easy computation to see that $\Erm \restrict \xi^\flat = u$. Then, for any $\eta \in \Wedge_2 \R^3$,
\[
  \eta \wedge (\Erm \restrict \xi^\flat)
  = \eta \wedge u
  = [\spr{\eta,\xi}\xi] \wedge u
  = \spr{\eta,\xi} \, \Erm,
\]
which shows that indeed $\Erm \restrict \xi^\flat = \hodge \xi$ by the definition of the Hodge star.

In conclusion, we have shown that in our situation the combined \term{dislocation current}
\[
  T := \sum_i q_i \otimes T_i
\]
is given as
\[
  T = b \otimes \frac{\dot\gamma}{\abs{\dot{\gamma}}} \; \Hcal^1 \restrict \im \gamma.
\]
In this way we have recovered the classical description of dislocations as $\R^3$-valued boundaryless integral $1$-currents, namely superpositions of edge and screw dislocations.

As the Burgers vector $b \in \Bcal$ is constant along $T$, we may instead work with the $1$-current
\[
  T^b := \vec{T}^b \, \Hcal^1 \restrict \im \gamma,  \qquad \vec{T}^b := \frac{\dot\gamma}{\abs{\dot{\gamma}}},
\]
for which it still holds that
\[
  \partial T^b = 0.
\]
In this way, $b$ may be considered the \enquote{topological charge} of $T^b$. If we generalize the above derivation and now collect \emph{all} dislocations with Burgers vector $b$ in $T^b$, we arrive at the description of dislocations used in the main part of this work. The compound current $T$ above can be computed from our representation $(T^b)_{b \in \Bcal}$ via
\[
  T = \int b \otimes \vec{T}^b \, \tv{T^b} \dd \kappa(b)
\]
In this sense, $T$ can be seen as (a singular version of) \term{Kr\"oner's dislocation density tensor $\alpha$}~\cite{Kroner01,Kroner60,Kroner61}. In the discrete case, where the $T^b$ are \emph{integral} $1$-currents (as in the present work), this representation is equivalent to ours because integral $1$-currents can only intersect in $\Hcal^1$-negligible sets (e.g., points). However, for \emph{fields} of loops, which can be represented by normal currents, $(T^b)_{b \in \Bcal}$ clearly contains more information than $T$.

Finally, we can unroll the above definitions to write
\[
  \int b \otimes \vec{T}^b \, \tv{T^b} \dd \kappa(b) = T = \sum_i q_i \otimes T_i = \sum_i q_i \otimes (\dbr{\R^3} \restrict d\vartheta^i),
\]
where $\dbr{\R^3} = \Erm \, \Lcal^3$ denotes the $3$-current associated with (integration over) $\R^3$. In referential coordinates, $\vartheta^i = \ee_i^T P = P^i$ (the $i$th row of $P$), where $P = Q^{-1}$ and $Q$ contains the vectors $q_1,q_2,q_3$ as columns. Thus, $d \vartheta^i$ takes the role of $\curl P^i$ (in the usual sense) and the above formula can be interpreted as a version of the consistency condition~\eqref{eq:consist} (see also Section~\ref{sc:consistproof}).

\subsection{Energy density and implants} \label{ax:implant}

We now posit that our material is \term{hyperelastic} in the sense of this term as used in the geometric paradigm: Assume that we are given an energy density $\hat{W} \colon \Trm^* M \otimes \R^3 \to \R$, i.e., $\hat{W}$ is a smooth section of $(\Trm^* M \otimes \R^3)^*$. Our connection $\nabla$ is called a \term{material connection} if its parallel transport operator $\Pi$ leaves $\hat{W}$ invariant: For all $x,x' \in M$, all $A \in \Trm^*_x M \otimes \R^3$, and all paths $\gamma$ from $x'$ to $x$ with associated (path-independent) parallel transport $\Pi_{x'}^x \colon \Trm_{x'} M \to \Trm_x M$, it needs to hold that
\begin{equation} \label{eq:Wp}
  \hat{W}_{x'}(A \circ \Pi_{x'}^x) = \hat{W}_x(A).
\end{equation}
For $x \in M$ define $Q_x := (q_1|_x,q_2|_x,q_3|_x) \colon \R^3 \to \Trm_x M$ (or, equivalently, $Q_x \in \Trm_x M \otimes \R^3$) to be the matrix representation of the basis $\{q_i|_x\}$. We have
\[
  Q_x = \Pi_{x'}^x \circ Q_{x'}.
\]
Define the \term{archetype} $W \colon \R^3 \otimes \R^3 \to \R$ via
\[
  W(F) := \hat{W}_x(F \circ Q_x^{-1}),  \qquad F \in \R^3 \otimes \R^3.
\]
Then, for $A \in \Trm^*_x M \otimes \R^3$, via~\eqref{eq:Wp},
\[
  \hat{W}_x(A)
  = \hat{W}_{x'}(A \circ \Pi_{x'}^x)
  = W(A \circ \Pi_{x'}^x \circ Q_{x'})
  = W(A \circ Q_x).
\]
In the terminology of~\cite{EpsteinKupfermanMaor20}, $Q_x$ is the \term{implant map} of $W$ at $x$. The expression of $\hat{W}_x$ via the archetype $W$ and the crystal scaffold frame $Q_x$ expresses the material uniformity, i.e., the material response is \enquote{the same} everywhere once the implant map is taken into account.

The problem of optimally embedding the body manifold $(M,g,\nabla)$ into an ambient Euclidean $\R^3$ then reads as follows: Find $y \colon M \to \R^3$ such that
\[
  \int_M \hat{W}_x(dy_x) \dd \mathrm{Vol}_g \to \min.
\]
This is now evidently equivalent to
\[
  \int_M W(dy_x \circ Q_x) \dd \mathrm{Vol}_g \to \min.
\]
This is exactly our elastic minimization problem (if $M = \R^3$ or a subset thereof). It can be shown (see~\cite[Section~5]{EpsteinKupfermanMaor20}) that in the \emph{isotropic} situation this expression in fact only depends on $g$, not on the particular choice of frame, so the energy integral is gauge-invariant with respect to the choice of basis.

\subsection{Translational gauge theories}

Finally, we note the translational gauge theory approach~\cite{Kleinert89book2,Lazar00,Katanaev05,Malyshev07,HehlObukhov07,LazarAnastassiadis08,LazarAnastassiadis09} is in a sense similar to the geometric paradigm, but now the coframe $\{\vartheta^i|_x\}_i$ is replaced by an Ehresmann connection $\vartheta$ in a principal $T(3)$-bundle, where $T(3)$ is the abelian Lie group of translations. In particular, since $T(3)$ is abelian, $\vartheta$ has field strength given by $d\vartheta$ (there is no homogeneous term). Then, the above arguments yield again the same description on the level of coordinates.  We refer to~\cite{HehlObukhov07} for a more detailed explanation of how torsion is related to a $T(3)$-gauge theory.


\providecommand{\bysame}{\leavevmode\hbox to3em{\hrulefill}\thinspace}
\providecommand{\MR}{\relax\ifhmode\unskip\space\fi MR }
\providecommand{\MRhref}[2]{%
  \href{http://www.ams.org/mathscinet-getitem?mr=#1}{#2}
}
\providecommand{\href}[2]{#2}

\end{document}